\documentclass[review]{elsarticle}
\usepackage{amsmath}
\usepackage{amsfonts,amssymb}
 \usepackage{lineno,hyperref}
\usepackage{graphicx}
\usepackage{subfig}
\usepackage{algorithm}
\usepackage{tabularx}
\usepackage{arydshln,leftidx,mathtools}
\usepackage{accents}
\usepackage[noend]{algpseudocode}
\journal{Journal of \LaTeX\ Templates}
\newcommand*{\dt}[1]{%
   \accentset{\mbox{\large\bfseries .}}{#1}}

\begin{document}

\begin{frontmatter}
\tableofcontents
\title{Can a patchy model describe the potential spread of  West Nile virus  in Germany? }
\author[2]{Suman Bhowmick\fnref{myfootnote}}
\author[1]{J\"orn Gethmann}
\author[2,3]{Igor M. Sokolov}
\author[1]{Franz J. Conraths}
\author[1]{Hartmut H. K. Lentz}

\address[1]{Friedrich-Loeffler-Institut\\ Institute of Epidemiology\\S\"udufer 10, 17493 Greifswald, Germany}
\address[2]{Institute of Physics, Humboldt University of Berlin\\Newtonstra{\ss}e 15, 12489 Berlin}
\address[3]{IRIS Adlershof, Zum Gro{\ss}en Windkanal 6, 12489 Berlin, Germany}
\fntext[myfootnote]{Corresponding author}




\begin{abstract}
In $2018$, West Nile Virus (WNV) was detected for the first time in Germany. 
Since the first detection, 36 human cases and 175 cases in horses and birds are detected. 
The transmission cycle of West Nile Virus includes birds and mosquitoes and – as dead-end hosts – people and horses. 
Spatial dissemination of the disease is caused by the movements of birds and mosquitoes. 
While the activity and movement of mosquitoes are depending mainly on temperature, in the birds there is a complex movement pattern caused by local birds and long range dispersal  birds.
To this end, we have developed a metapopulation network model framework to delineate the potential spatial distribution and spread of WNV across Germany as well as to evaluate the risk throughout our proposed network model.
Our model facilitates the interconnection amongst the vector, local birds and long range dispersal birds contact networks. 
We have assumed different distance dispersal kernels models for the vector and avian populations with the intention to include short and long range dispersal.
The model includes spatial variation of mosquito abundance and the movements to resemble the reality.

\end{abstract}


\end{frontmatter}


\section{Introduction}

In August $2018$, West Nile Virus (WNV), was detected for the first time in Germany \cite{ZIEGLER201939}. 
Genetic characterisation indicates, that the German cases are in the same Central European subclade as cases found in the Czech Republic and Austria \cite{ZIEGLER201939}. 
Since the first detection, 36 human cases and 175 cases in horses and birds are detected. 
The disease pattern is mainly focussed on the eastern part of Germany.
The disease is maintained by an enzootic transmission cycle between birds and mosquitoes. 
Humans and horses can get infected but they will not spread the disease \cite{Chancey.2015}.
The spread of mosquito-borne infectious diseases is a spatio-temporal dynamic process that is being affected by multiple agents such as vector and host movements, pathogen transmission heterogeneity, environmental factors etc~\cite{doi:10.1111/1365-2435.12645, White7374}. 
Previous studies have revealed that increase in the temperature and the host-vector mobility facilitate the spread of mosquito-borne diseases~\cite{10.1093/icb/icw015,10.1371/journal.pbio.2003489}.
According to~\cite{ZIEGLER201939}, the summer of $2018$ have provided the favourable climatic conditions for the potential geographical spread of zoonotic arthropod-borne WNV in Germany and possibly it have been introduced by the wild birds as they can act as amplifying hosts.
A previous study on the WNV outbreak in the United States \cite{Kilpatrick2007} has listed the likely pathways by which WNV has spread as migratory birds, dispersal of nonmigratory birds or long range birds, movement of mosquitoes by flight or wind, and human transport of mosquitoes, birds, or other animals. 
While \cite{Rappole, Reed} state out that migratory or the long range dispersal birds may play the significant role as a spreader to new regions along their major flyways across the globe, \cite{Kilpatrick2007} concludes that despite the fact that many studies have carried out, there is still no evidence for this hypothesis.

In general, there are two migrating seasons for the long dispersal birds, spring and autumn \cite{newton2010migration}. 
In the spring migration season birds are migrating from the south northwards to Germany for breeding. 
Once the birds arrived in Germany, they will breed and stay for summer in the same local area. 
In autumn, they leave Germany and migrate into the south. 
On their way from the south, they might get infected and introduce the virus into Germany. 
While they are in Germany, the temperature might be suitable for disease transmission  \cite{https://doi.org/10.1111/ele.13335, 10.1371/journal.pbio.3000938}. 

On the other hand, in autumn, birds migrating from the North and East to overwinter in Germany. 
These birds play a minor or no role in WNV transmission as temperature in winter is not suitable for transmission.

Another potential route of introduction are local birds, or vectors, spreading the virus on a local area.
In 2017, \cite{rudolf2017west} shows that they detect WNV positive vectors in the Czech republic, other cases were found in Hungary, and Austria \cite{v12010123, v11070639, https://doi.org/10.1111/tbed.13452}.
\par

Other studies \cite{doi:10.1098/rspb.2017.1807, Tatem6242} have included the importance of host movement to analyse the infection transmission through spatial host networks in the heterogeneous environment. 
In \cite{Sumner}, the authors quantify the importance of movement of livestock and the dispersal of vector in the disease transmission. 

\par
After the first introduction of WNV to birds and equines in Germany, the cases of WNV has increased in the following season \cite{Ute}.
The activity of WNV is detected in the Eastern part to the Northern zone of Germany.
The combination of phylogenetic analysis and the wide distribution of WNV in Germany from north to the south reveals that WNV may have been introduced to Germany from Czech Republic already before $2018$ \cite{ZIEGLER201939}.
These findings demonstrate that there is a further risk of potential spatial transmission of WNV in Germany in possible with some additional cases in the mammals and in the birds.

In our current work we are interested in the description of the spatial spread of WNV in Germany under the effect of host (bird) and vector (mosquito) movements, including seasonality. 
Once WNV is introduced into Germany, the main actors in the local and spatial transmission are local birds and vectors. 
In this study, we assume, that migratory or the long distance dispersal birds that settle down for breeding will have a similar behaviour as local birds.

\par

Geographical and population movements are essential in the context of spatial transmission.
There are several approaches to model the spatial transmission of geographical and population movements. 
Partial differential equation (PDE)\cite{MAIDANA2009403, Tar, Wang} are one of the choice for that.
However, while modelling the geo-spatial dissemination of disease, usually there is a separation between the diffusion and the dispersal models are being made.
In diffusion model, the transmission occurs immediately to the neighbouring zones but in patchy environment, this kind of modelling assumptions are not preferable.\cite{kenkre2005theoretical} described a general approach, how to model the spatial transmission of WNV by a diffusion model.
Moreover, given the extent of spatial spread of WNV across Germany possibly due to the migratory or long range dispersal birds, distance based dispersal models seem to be more applicable.
While accounting on the dispersal model, metapopulation model is a valuable modelling approach for such purposes.
In \cite{BICHARA2016128, COLIZZA2008450}, the authors investigate the impact of the host dispersal amongst the multiple patches in the disease dynamics.
Spatio-temporal features in the progress of infectious agents of different hosts are in included in \cite{Arino06diseasespread, doi:10.1142/9789814261265_0003} using metapopulation models.
In \cite{Arin1, Arin2}, the authors deal with the multi-species epidemic models on n patches with migration what can potentially be employed in the vector-hosts model.
\par
Most WNV spread models are mathematical deterministic compartmental models \cite{refId0, doi:10.1137/18M1236162, Gumel}.
However, these models are usually developed at a local scale that do not necessarily include the global information about the different components such as mobility patterns of the hosts or vector or both, temperature or landscape data types.
In a previous study, we develop a temperature driven model~\cite{BHOWMICK2020110117} that analyses the local spread of different regions in Germany. 
In this study, we extend this model to a metapopulation-network associated model.

In contrast to the results of our previous model~\cite{BHOWMICK2020110117}, where the most suitable region for the establishment of the disease is in the South West, the cases in Germany are mainly observed in the East. 

The new model should help us to understand the WNV spatial transmission in Germany, and to understand the key factors of spatio-temporal transmission of WNV.  
Hence, we systematically examine the relevance of the variables in our model.
Our current  endeavour is similar in spirit to other several models \cite{10.1371/journal.pcbi.1006875, refId1, LAPERRIERE201634} that also direct to bring necessary information in the spatial scales. 

\section{Model Formulation}

\begin{figure}[!h]
\centering
\subfloat[Metapopulation]{
  \includegraphics[height=35mm,width=60mm]{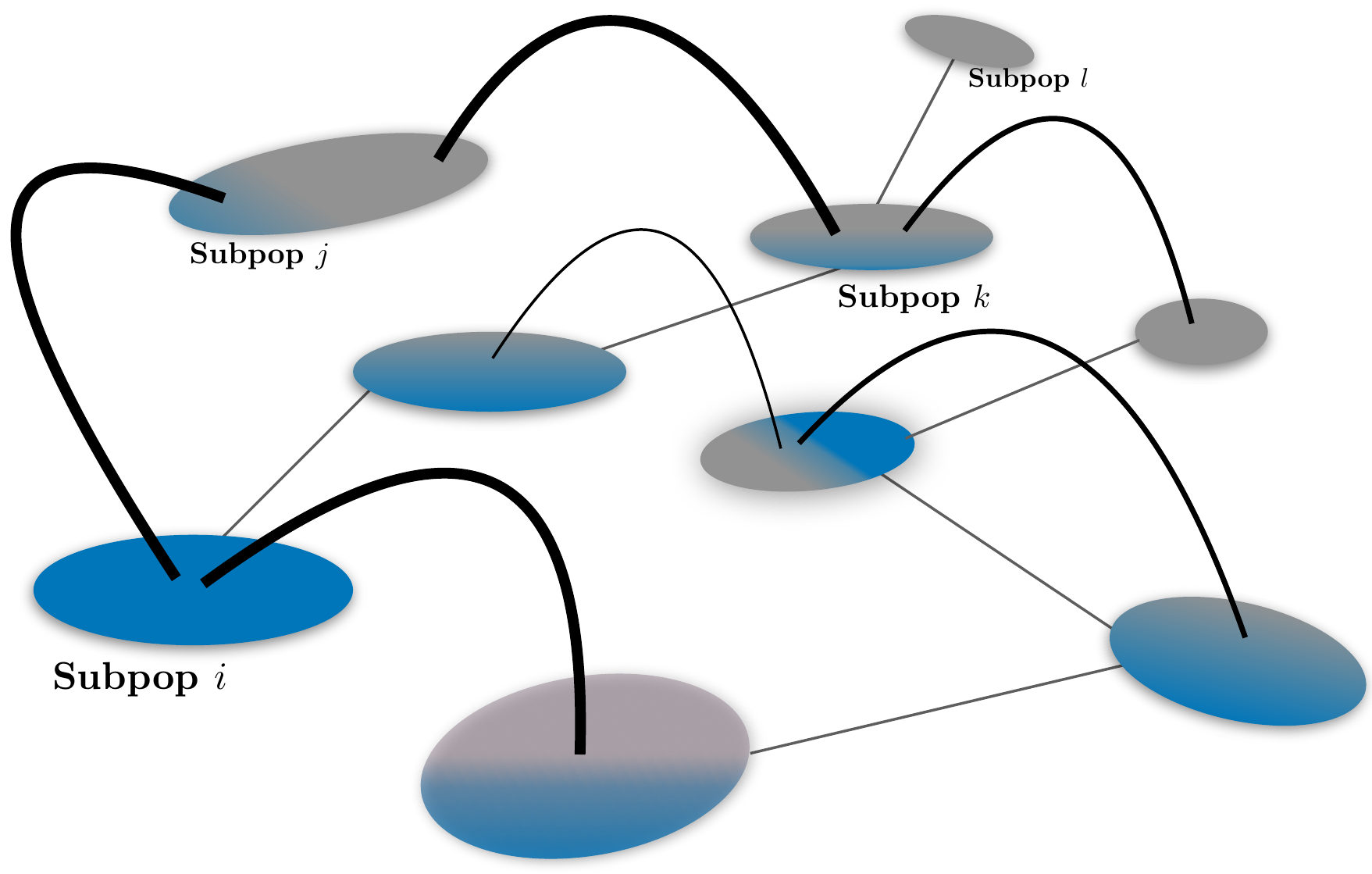}
}
\subfloat[Subpopulation \& Individuals]{
  \includegraphics[height=35mm,width=60mm]{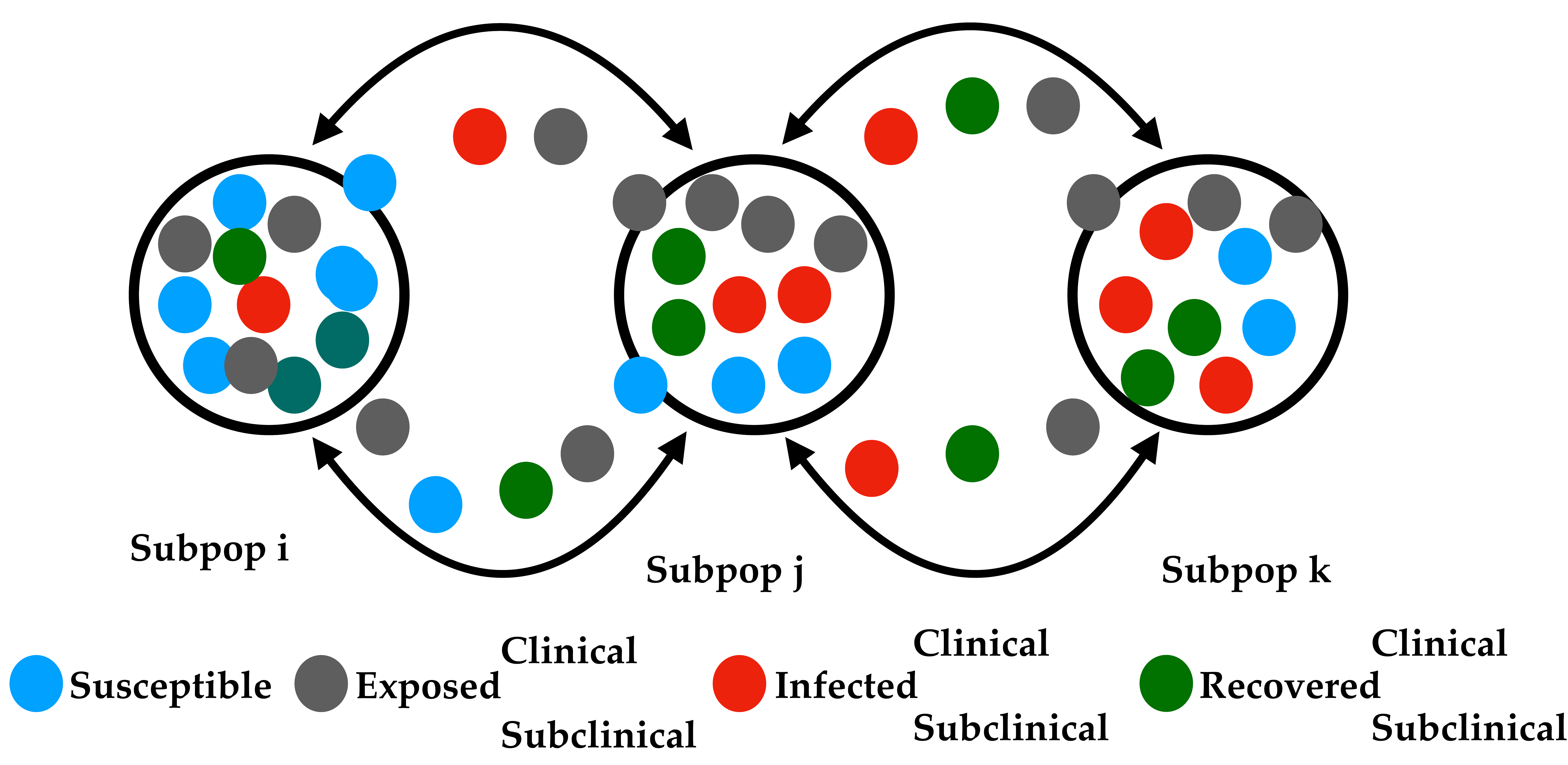}
} 
\caption{Schematic representation of a metapopulation model and its movements.
(a) Metapopulation: The figure indicates the different patches in a metapopulation model and their connections. 
(b) The figure illustrates the individuals of a compartment model moving between patches. 
Each patch consists of population of individuals who are represented with respect to their health status (e.g., susceptible, exposed, infected, removed) in different colours.
Exposed, Infected and the recovered classes are further subdivided into the clinical and the subclinical cases.
}
\label{Fig:Metapop1}
\end{figure}

Using our local model as a basis~\cite{BHOWMICK2020110117}, we have constructed a metapopulation model. 
The principle functionality of a metapopulation model is shown in figure \ref{Fig:Metapop1}. 
Our model system comprises of heterogeneous networks of subpopulations or patches what are connected by migration. 
Each subpopulation represents the population of vectors and hosts in the habitat patch.
The respective migrations of individuals from one subpopulation to another subpopulation is governed by the migration paths of connections among the subpopulations.
Individuals can migrate from a subpopulation to another on the network of connections among subpopulations. 
Transmission within each patch  is modelled by a vector-host compartment model.
Nonlinear Ordinary Differential Equations are employed to model the local transmission and the transmission due to the migration amongst the adjoining habitat sites as well the population dynamics.


\subsection{Migration Model}
Movement rate from a patch $i$ to another patch $j$ is denoted by $r_{ij} \geq 0$.
It is assumed that population growth, birth and the disease transmission happen in the respective patches only but not during the migration.
Assuming $Q_i$ is the population in patch $i$, then the dynamics associated with $Q_i(t)$ is given by for $i =1, ...,n$

\begin{equation} \label{Eq:1.a}
\frac{dQ_{i}}{dt} =  \Pi_{Q_{i}} - \mu_{Q_{i}}Q_{i}+\sum_{j} r_{ij}Q_{j}-\sum_{j} r_{ji}Q_{i}
\end{equation}
 where $\Pi_{Q_{i}}$ represents recruitment in the population $Q_i$,  $\mu_{Q_{i}}$ is the death rate at the patch $i$.
 Equation~\eqref{Eq:1.a} can be expressed in a more compact form as 
 
 \begin{equation} \label{Eq:1.b}
\dt{Q} = \Pi_{Q}- diag(\mu_{Q})Q+RQ
 \end{equation}
 
 where, $Q = (Q_1, Q_2, ..., Q_n)^\intercal$, $\Pi_{Q} = (\Pi_{Q_{1}}, \Pi_{Q_{2}},...,\Pi_{Q_{n}})^\intercal$, $\mu_{Q} = (\mu_{Q_{1}}, \mu_{Q_{2}},...,\mu_{Q_{n}})^\intercal$, and
 $R$ is the movement matrix and the elements of $R$ are defined as $R_{ij}= r_{ij}$ for $i \neq j$.

  \subsection{Epidemiological Models For Local Birds \&  Mosquitoes}
 
 When combining the local epidemiological model~\cite{BHOWMICK2020110117} and the metapopulation model subsequently will lead to equation \ref{Eq:1} representing the model for the mosquitoes with $m_{ij}$ as the rate of migration movement between two arbitrary nodes $j$ and $i$.
   \begin{eqnarray} \label{Eq:1}
   \frac{dS_{M_{i}}}{dt} &=&\left[b_{M_{i}}N_{M_{i}}-m_{M_{i}}S_{M_{i}}\right]\left[1-S_{M_{i}}/K_{M_{i}}\right]-\frac{S_{M_{i}}}{K_{B_{i}}}\left(c_{2_{i}}I_{BC_{i}}+c_{1_{i}}I_{BSC_{i}}\right) \nonumber  \\ 
   &+&\sum_{j} m_{ij}S_{M_{j}}-\sum_{j} m_{ji}S_{M_{i}}\nonumber  \\ 
\frac{dE_{M_{i}}}{dt} &=&   \frac{S_{M_{i}}}{K_{B_{i}}}\left(c_{2_{i}}I_{BC_{i}}+c_{1_{i}}I_{BSC_{i}}\right)-\gamma_M E_{M_{i}}-m_{M_{i}} E_{M_{i}}+\sum_{j} m_{ij}E_{M_{j}}-\sum_{j} m_{ji}E_{M_{i}}\nonumber  \\ 
\frac{dI_{M_{i}}}{dt} &=& \gamma_M E_{M_{i}}-m_{M_{i}} I_{M_{i}}+\sum_{j} m_{ij}I_{M_{j}}-\sum_{j} m_{ji}I_{M_{i}}
       \end{eqnarray}
  
Equation \ref{Eq:2} shows the model for birds and $p_{ij}$ represents the migration between two arbitrary nodes $j$ \& $i$ of the birds.

 \begin{eqnarray} \label{Eq:2}
\frac{dS_{B_{i}}}{dt} &=& \left[ b_{B_{i}}-(b_{B_{i}}-m_{B_{i}}) \frac{N_{B_{i}}}{K_{B_{i}}}\right]-m_{B_{i}}S_{B_{i}} - \frac{(\beta_{3_{i}}+\beta_{4_{i}})I_{M_{i}}S_{B_{i}}}{K_{B_{i}}}\nonumber  \\ 
   &+&\sum_{j} p_{ij}S_{B_{j}}-\sum_{j} p_{ji}S_{B_{i}}\nonumber  \\
\frac{dE_{BC_{i}}}{dt} &=& \frac{\beta_{4_{i}}I_{M_{i}}S_{B_{i}}}{K_{B_{i}}}-m_{B_{i}}E_{BC_{i}}-\gamma_{BC_{i}}E_{BC_{i}}+\sum_{j} p_{ij}E_{BC_{j}}-\sum_{j} p_{ji}E_{BC_{i}}\nonumber  \\
\frac{dE_{BSC_{i}}}{dt} &=& \frac{\beta_{3_{i}}I_{M_{i}}S_{B_{i}}}{K_{B_{i}}}-m_{B_{i}}E_{BSC_{i}}-\gamma_{BSC_{i}}E_{BSC_{i}}+\sum_{j} p_{ij}E_{BSC_{j}}-\sum_{j} p_{ji}E_{BSC_{i}}\nonumber  \\
\frac{dI_{BC_{i}}}{dt} &=& \gamma_{BC_{i}}E_{BC_{i}}-m_{B_{i}}I_{BC_{i}}-\alpha_{4_{i}}I_{BC_{i}}-d_{BC_{i}}I_{BC_{i}}+\sum_{j} p_{ij}I_{BC_{j}}-\sum_{j} p_{ji}I_{BC_{i}}\nonumber  \\
\frac{dI_{BSC_{i}}}{dt} &=& \gamma_{BSC_{i}}E_{BSC_{i}}-m_{B_{i}}I_{BSC_{i}}-\alpha_{3_{i}}I_{BSC_{i}}+\gamma_{3_{i}}R_{BSC_{i}}\nonumber \\
&-&d_{BSC_{i}}I_{BSC_{i}}+\sum_{j} p_{ij}I_{BSC_{j}}-\sum_{j} p_{ji}I_{BSC_{i}}\nonumber  \\
\frac{dR_{BC_{i}}}{dt} &=& \alpha_{4_{i}}I_{BC_{i}}-m_{B_{i}}R_{BC_{i}}+\sum_{j} p_{ij}R_{BC_{j}}-\sum_{j} p_{ji}R_{BC_{i}}\nonumber  \\
\frac{dR_{BSC_{i}}}{dt} &=& \alpha_{3_{i}}I_{BSC_{i}}-m_{B_{i}}R_{BSC_{i}}-\gamma_{3}R_{BSC_{i}}+\sum_{j} p_{ij}R_{BC_{j}}-\sum_{j} p_{ji}R_{BC_{i}}
 \end{eqnarray}

Similar expressions stand for the long range bird population but we decide not put that here.
We have appended in the Supplementary Information. 
The initial conditions are\\ 
$S_{B_{i}}(0), S_{M_{i}}(0) > 0 \, \& \\ E_{BC_{i}}(0), E_{BSC_{i}}(0), I_{BC_{i}}(0), I_{BSC_{i}}(0), R_{BC_{i}}(0), R_{BSC_{i}}(0), E_{M_{i}}(0), I_{M_{i}}(0)\geq0$.
\par

Let $\lambda_i = \frac{\beta_{3_{i}} I_{M_{i}}} {K_{B_{i}}} $, $\eta_i = \frac{\beta_{4_{i} }I_{M_{i}}}{K_{B_{i}}} $, $\delta_i = \frac{c_{2_{i}}I_{BC_{i}}}{K_{B_{i}}}$ and $\mu_i = \frac{c_{1_{i}}I_{BSC_{i}}}{K_{B_{i}}}$.\\
$ \Lambda_i = \left[b_{M_{i}}N_{M_{i}}-m_{M_{i}}S_{M_{i}}\right]\left[1-S_{M_{i}}/K_{M_{i}}\right], \Pi_i = \left[ b_{B_{i}}-(b_{B_{i}}-m_{B_{i}}) \frac{N_{B_{i}}}{K_{B_{i}}}\right]$

Adding up Eqn~\eqref{Eq:1} \,\, \& Eqn~\eqref{Eq:2}  give us equations for the total mosquito and birds populations, respectively, in patch $i = 1, . . . , n:$ while using the notations introduced above,
we get 
\begin{eqnarray} \label{Eq:2a}
\frac{dN_{B_{i}}}{dt} &=&  \Pi_i -m_{B_{i}}N_{B_{i}}-d_{BC_{i}}I_{BC_{i}}-d_{BSC_{i}}I_{BSC_{i}}\nonumber  \\
                               &+& \sum_{Z}\left(\sum_{j} p_{ij}Z_{j}-\sum_{j} p_{ji}Z_{i}\right)\\
\frac{dN_{M_{i}}}{dt} &=& \Lambda_i -m_{M_{i}}N_{M{i}}+ \sum_{Y}\left(\sum_{j} m_{ij}Y_{j}-\sum_{j} m_{ji}Y_{i}\right)\                              
 \end{eqnarray}
Here, $Y = S_M, E_M, I_M$ and  $Z = S_B, E_{BC}, E_{BSC}, I_{BC}, I_{BSC}, R_{BC}, R_{BSC}$.
Let the total bird and mosquito population be denoted as $N_B$ and $N_M$, respectively.
Then after adding the population over all the patches, we should get the following
\begin{equation} \label{Eq:2b}
\frac{dN_B}{dt} = \sum_{i}\left(\Pi_i-m_{B_{i}}N_{B_{i}}-d_{BC_{i}}I_{BC_{i}}-d_{BSC_{i}}I_{BSC_{i}}\right)+\sum_{i}\left[\sum_{Z}\left(\sum_{j} p_{ij}Z_{j}-\sum_{j} p_{ji}Z_{i}\right)\right]
\end{equation}
Since $I_{BC_{i}} < N_{B_{i}}$ \& $I_{BSC_{i}} < N_{B_{i}}$, it follows that
\begin{equation} \label{Eq:2c}
\sum_{i}\Pi_i -\sum_{i} (m_{B_{i}}+d_{BC_{i}}+d_{BSC_{i}})N_{B_{i}} \leq \frac{N_{B_{i}}}{dt} \leq \sum_{i}\Pi_i-\sum_{i} m_{B_{i}}N_{B_{i}}
\end{equation}

Thus
\begin{equation} \label{Eq:2d}
\sum_{i}\Pi_i -\max_{1\leq i \leq n}\{m_{B_{i}}+d_{BC_{i}}+d_{BSC_{i}}\}N_{B_{i}}\leq \frac{N_{B_{i}}}{dt} \leq \sum_{i}\Pi_i \min_{1\leq i \leq n}\{m_{B_{i}}\}N_{B_{i}}
\end{equation}
From here, we can claim that $\forall t \geq 0$,
\begin{equation} \label{Eq:2e}
\min\left[\frac{\sum_{i}\Pi_i}{\max_{1\leq i \leq n}\{m_{B_{i}}+d_{BC_{i}}+d_{BSC_{i}}\}}, N_B(0)\right] \leq N_B(t) \leq \max\left[\frac{\sum_{i}\Pi_i}{\max_{1\leq i \leq n}\{m_{B_{i}}\}}, N_B(0)\right]
\end{equation}
Therefore, the total populations of local birds are bounded.

We can reframe  ~\ref{Eq:1} \&  ~\ref{Eq:2} into the matrix form as follow:
 \begin{eqnarray} \label{Eq:3}
 \dt{S_{M}} &=& \Lambda -diag(\delta + \mu ) S_M + MS_M \nonumber  \\
\dt{E_{M}} &=& diag(\delta + \mu ) S_M -diag(\gamma_M + m_M ) E_M+ ME_M \nonumber  \\
\dt{I_{M}} &=& diag(\gamma_M ) E_M -diag(m_M) I_M+ MI_M \nonumber  \\
 \end{eqnarray}
And for the local birds:
\begin{eqnarray} \label{Eq:4}
 \dt{S_{B}} &=& \Pi -diag(\lambda + \eta +m_B) S_B + PS_B \nonumber  \\
 \dt{E_{BC}} &=& diag(\eta ) S_B -diag(\gamma_{BC} + m_B ) E_{BC}+ PE_{BC} \nonumber  \\
 \dt{E_{BSC}} &=& diag(\lambda ) S_B -diag(\gamma_{BSC} + m_B ) E_{BSC}+ PE_{BSC} \nonumber  \\
\dt{I_{BC}} &=& diag( \gamma_{BC}) E_{BC} -diag(\alpha_4 + m_B +d_{BC}) I_{BC}+ PI_{BC} \nonumber  \\
 \dt{I_{BSC}} &=& diag( \gamma_{BSC}) E_{BSC} -diag(\alpha_3 + m_B+d_{BSC} ) I_{BSC}+ diag(\gamma_3)R_{BSC}+PI_{BSC} \nonumber \\
 \dt{R_{BC}} &=&diag(\alpha_4)I_{BC}-diag(m_B)R_{BC}+PR_{BC} \nonumber \\
 \dt{R_{BSC}} &=&diag(\alpha_3)I_{BSC}-diag(m_B+\gamma_3)R_{BSC}+PR_{BSC},    
 \end{eqnarray}
where $M$, $P$ are the movement matrices of the vector and the local bird population, respectively.
The matrix form of the equations associated with the long range dispersal  birds are included in the Supplementary Information.

\section{Network Framework}
In the last section we show the ODE model for WNV disease spread between patches.
In this section we describe the mobility networks of mosquitoes and birds. 
These networks are necessary to define the movement matrices for the vector ($M$) and the local bird ($P$) populations.

\subsection{Vector Mobility Network}\label{Vector Mobility Network}
The importance of vector movement in the spread of WNV is controversial. 
While ~\cite{MAIDANA2009403}  conclude that mosquito movements do not play an important role but ~\cite{ doi:10.1086/521911, Styer2007, Vogels2017, 10.1371/journal.pntd.0002768} conclude that due to the opportunistic bites from the mosquitoes to the host, mosquito movements are important.
Information on exact mobility pathways for mosquitoes are scarce.  
However, there are some estimations on the flight range of Culex ~\cite{doi:10.1093/jmedent/31.3.508, VERDONSCHOT201469, Cluexvol}. 
\cite{VERDONSCHOT201469} estimate the flight range of \textit{Culex pipiens} ($average\:maximum\:distance = 9695\:m$, $minimum\:of\:maximum\:distance  = 350\:m$, and $maximum\:of\:maximum\:distance = 22,530\:m$) and the dispersal capacity is \textit{ Strong}. 
Given their dispersal capacity, it is more realistic to include the precise and daily movement of the mosquitos. 
Later on we  show that the mosquito movement matters in case of spreading the disease from one patch $i$ to the neighbouring patch $j$.
Let the distance between two patches be ($i \&  j$) as $D_{ij}$, then according to  ~\cite{doi:10.1002/eap.1612},  the dispersal rates between two sub-populations ($M_{i,j}$) are assumed to be negative-exponential distribution. But for the simplicity we decided to follow the distribution proposed by \cite{MOULAY2013129}. 
Here, we have used the fact that Culex pipens have \textit{ Strong} dispersal capability, henceforth the $D_{max}$ used by ~\cite{MOULAY2013129} is different than what we have considered ~\cite{VERDONSCHOT201469} but the dispersal probability is a function of the linear decreasing distance as in ~\cite{MOULAY2013129}.
The network with such linear dispersal kernel is calculated as follows:
\begin{algorithm}
\caption{Mosquito movement network algorithm}\label{Alg:MosNet}
\begin{algorithmic}[1]
\Procedure{MosNet}{$i,j$}\Comment{Routine to create link between i \& j}
\State $D_{ij}\gets Euclidean \_ Distance (i, j)$ \Comment{Euclidean distance between i \& j}
\State $D_{max}\gets Maximum \_ Dist$ \Comment{Maximum interaction radius of mosquitoes}
\State $p(D_{ij})\gets \frac{D_{max}-D_{ij}}{D_{max}}$ \Comment{Probability of link connection between i \& j}
\State $p_{rand}\gets rand(0,1)$ \Comment{Generate a random  no between 0 \& 1}
\If {$p_{rand} < p(D_{ij})$}
\State Create an undirected link between $i  \&  j$ 
\EndIf
\EndProcedure
\end{algorithmic}
\end{algorithm} 

\begin{figure}[h!]
    \centering
     \includegraphics[width=40mm]{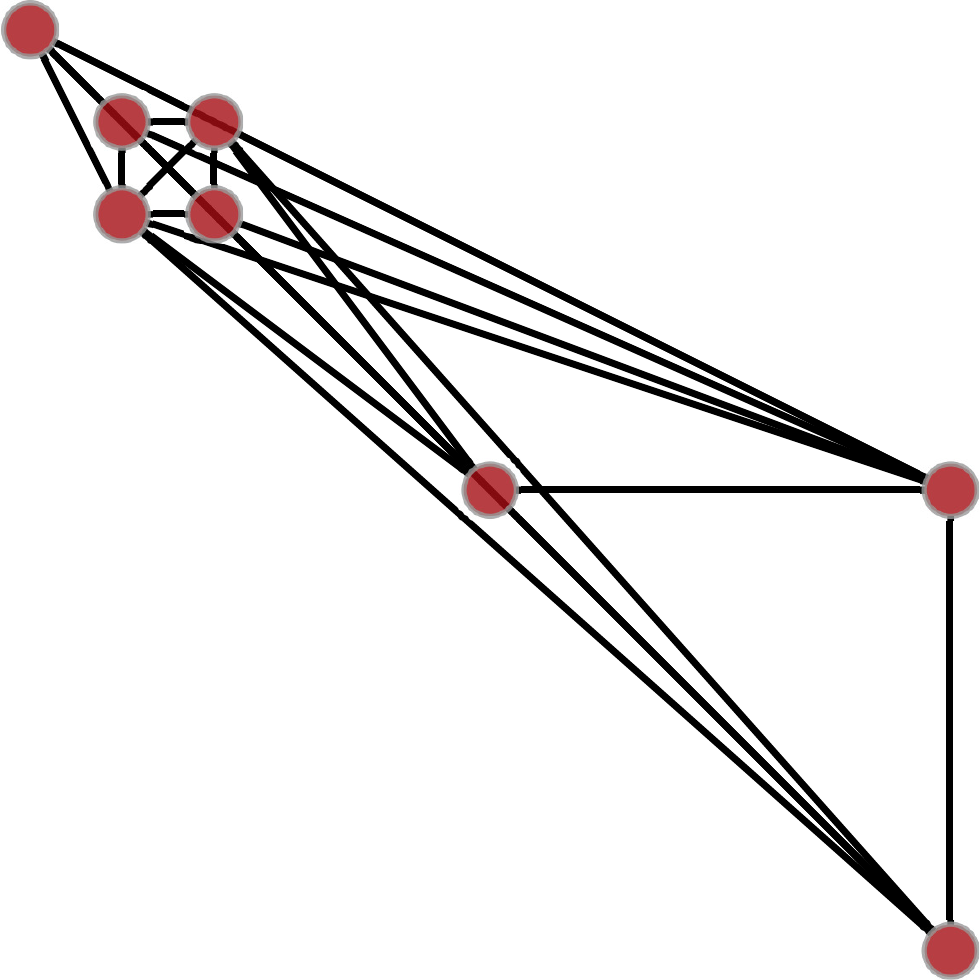}\\
    \caption{One realisation of the stochastic network of Mosquito}
    \label{Fig:MosPop}
\end{figure}

In figure  ~\ref{Fig:MosPop} we show an example of a randomly generated mosquito network  

\subsection{Host Mobility Model}
\begin{figure}[h!]
    \centering
     \includegraphics[width=50mm]{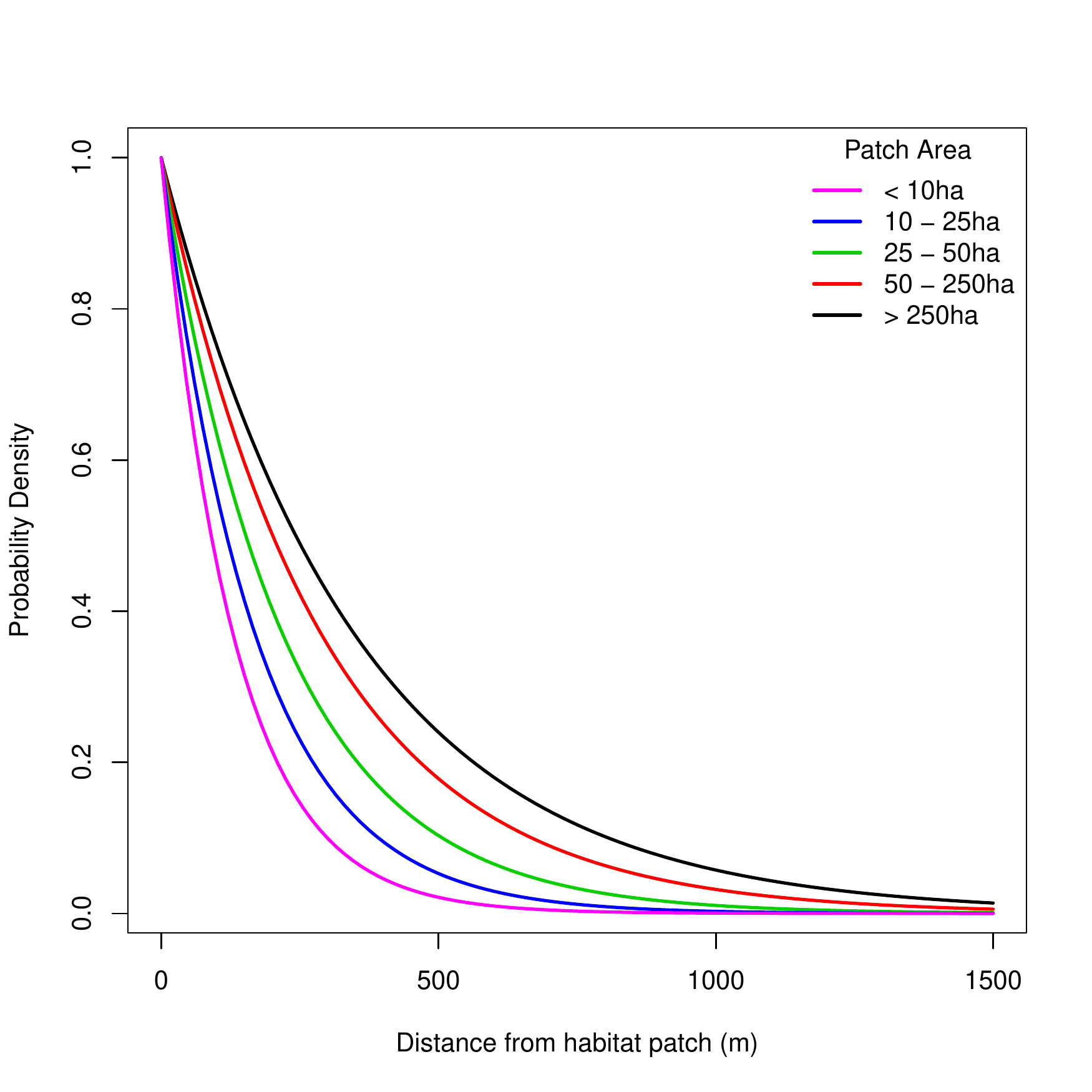}\\
    \caption{Dispersal probability of the bird}
    \label{Fig:LocalBird}
\end{figure}

Birds are the natural reservoir for WNV \cite{campbell2002west, Kilpatrick2007, liu2006modeling}. 
In general, there is a transmission cycle between birds and vectors. 
Furthermore, there are discussions about the importance of bird-to-bird transmissions \cite{hartemink2007importance} but in our current endeavour we have not included such transmission routes. 
It is obvious, that bird movement might spread the disease within the home range of birds. 
The home range of birds depends on the required habitat, as well as feed supply and bird density ~\cite{8937990, Butler2018, ref2018-32206-003}.  
As home ranges differs between species and depends on habitat suitability, it is difficult to define a single home range for the host mobility model.
Hence, we include two movement patterns, one to cover small home ranges of breeding birds with a maximal dispersal ranges from $1500$ to $2164$ meter and minimal dispersal ranges from $80$ to $170$ meter \cite{https://doi.org/10.1111/1365-2745.12685}
and a second one to cover large home ranges with a maximum flight distance of $500$ Km after following \cite{10.1371/journal.pcbi.1006875}.
In the Figure: \ref{Fig:LocalBird}, we show an example for the local bird flight range. 

\subsubsection{Local Bird Mobility Model} \label{LocalBird}

But, given the above said reasons, we decide to follow the seed dispersal model ~\cite{10.1371/journal.pone.0156688,trove.nla.gov.au/work/217129952, doi:10.1111/j.1365-2745.2008.01401.x, Banos-Villalba2017} for the local bird mobility model. 
We use the seed dispersal as the proxy for the movement network of the birds.
 The dispersal probability is of Weibull distribution ~\cite{doi:10.1111/j.1365-2745.2008.01379.x, NATHAN2008638, doi:10.1111/j.1469-8137.2011.04051.x, doi:10.1111/j.0906-7590.2006.04677.x, doi:10.1890/15-0734.1, 10.1371/journal.pone.0193660, SeedDispersalMapper}.

To construct the movement matrix ($P$) of the local birds, we have made use of the following routine just the way we have constructed the mosquito movement network ($M$) in  Algorithm:~\ref{Alg:MosNet} .
\begin{algorithm}
\caption{Local bird movement network algorithm}\label{Alg:BirdNet}
\begin{algorithmic}[1]
\Procedure{LocalBirdNet}{$i,j$}\Comment{Routine to create link between i \& j}
\State $D_{ij}\gets Haversine \_ Dist (i, j)$ \Comment{Haversine distance between i \& j}
\State $p(D_{ij})\gets Weibull \_ Distribution (D_{i j})$ \Comment{Probability of link connection between i \& j}
\State $p_{rand}\gets rand(0,1)$ \Comment{Generate a random  no between 0 \& 1}
\If {$p_{rand} < p(D_{ij})$}
\State Create an undirected link between $i  \&  j$ 
\EndIf
\EndProcedure
\end{algorithmic}
\end{algorithm}

\begin{figure}[!th]
    \centering
     \includegraphics[width=30mm]{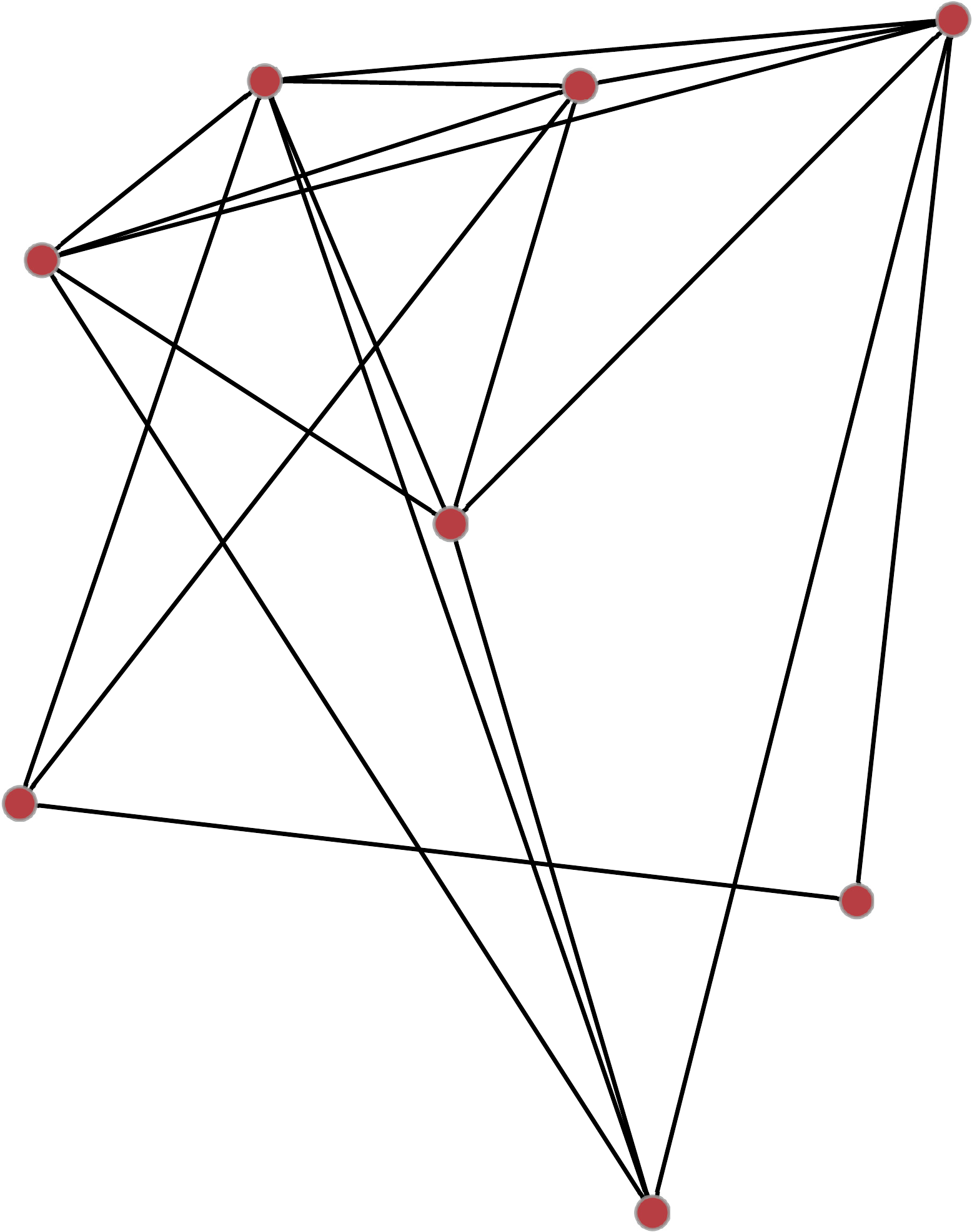}\\
    \caption{This is one realisation of the stochastic network of local bird}
    \label{Fig:BirdPop}
\end{figure}

An example of such generated local bird network using  Algorithm:~\ref{Alg:BirdNet} is shown in Figure: ~\ref{Fig:BirdPop} 
Shape and the scale parameters are taken from \cite{https://doi.org/10.1111/1365-2745.12685, 10.1093/aobpla/plz042, https://doi.org/10.1111/j.1365-2745.2008.01401.x} and the ranges for the shape and scale parameters are following  $[2.83, 3.26]$ and  $[0·01, 1.0]$.

\subsubsection{Long Range Dispersal Bird Mobility Model}

The information for large home ranges in birds are scarce. 
Hence, we follow the approach of~\cite{doi:10.1080/00107510500052444, 10.1371/journal.pcbi.1006875}, using a power-law distribution to model incidental long-range disease transmission routes. 
 To model the spatial dynamics of different infectious diseases, power-law transmission is used in several occasions ~\cite{meyer2014}. 
We have constructed the movement matrix ($N$) of the long dispersal bird using the following routine as we have done for the mosquito movement network and the local bird movement network in  
Algorithm:~\ref{Alg:MosNet} and Algorithm:~\ref{Alg:BirdNet}.


\begin{algorithm}
\caption{Long range dispersal bird movement network algorithm}\label{Alg:MigBirdNet}
\begin{algorithmic}[1]
\Procedure{MigBirdNet}{$i,j$}\Comment{Routine to create link between i \& j}
\State $D_{ij}\gets Haversine \_ Dist (i, j)$ \Comment{Haversine distance between i \& j}
\State $p(D_{ij})\gets Powerlaw \_ Distribution (D_{i, j})$ \Comment{Probability of link connection between i \& j}
\State $p_{rand}\gets rand(0,1)$ \Comment{Generate a random  no between 0 \& 1}
\If {$p_{rand} < p(D_{ij})$}
\State Create an undirected link between $i  \&  j$ 
\EndIf
\EndProcedure
\end{algorithmic}
\end{algorithm}

\begin{figure}[!th]
    \centering
     \includegraphics[width=30mm]{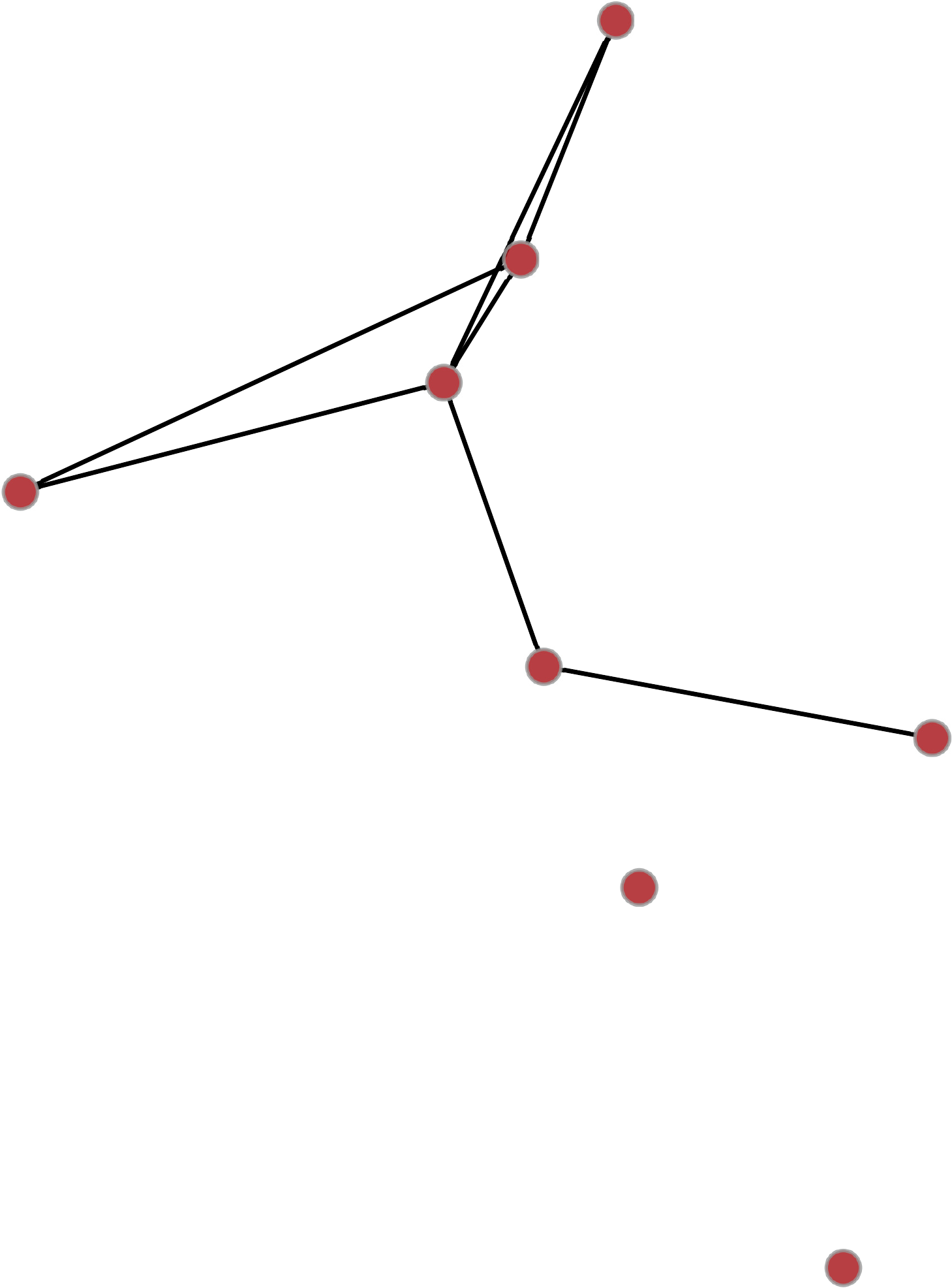}\\
    \caption{One realisation of the stochastic network of long range dispersal bird}
    \label{Fig:MigBirdPop}
\end{figure}

An example of such generated long range dispersal bird network using  Algorithm:~\ref{Alg:MigBirdNet} is shown in Figure: ~\ref{Fig:MigBirdPop} 

The value of the power-law parameter has been taken from  \cite{10.1371/journal.pcbi.1006875} with its value ranging $\sim$ $U(2,4)$. 

\subsection{Mathematical Preliminaries}
\subsubsection{Disease Free Equilibrium}
To find the disease free equilibrium $E_0$, we consider the following linear system
\begin{eqnarray} \label{Eq:5}
\Lambda -diag(\delta + \mu ) S_M + MS_M&=& 0 \nonumber  \\
\Pi -diag(\lambda + \eta +m_B) S_B + PS_B&=& 0
\end{eqnarray}   
or, in the compact form as
\begin{eqnarray} \label{Eq:5a}
HS = \Omega
\end{eqnarray}
where 
 \[
H=
  \begin{bmatrix}
    diag(\delta+\mu )-M &0 \\
    0 & diag(\lambda + \eta +m_B)-P
  \end{bmatrix},  
S =
  \begin{bmatrix}
    S_M \\
    S_B
  \end{bmatrix}
  \Omega =
  \begin{bmatrix}
    \Lambda \\
    \Pi
  \end{bmatrix}
  \]
Since all off-diagonal entries of $H$ are nonpositive and the sum of the entries in each column of $H$ is positive, $H$ is a nonsingular $M$-matrix, $H^{-1}\geq 0$~\cite{Abraham}.
Therefore, the linear system Eqn\eqref{Eq:5a} has a unique positive solution 
$S^0 = (S_{M_{1}}^{0}, S_{M_{2}}^{0}, ..., S_{M_{n}}^{0}, S_{B_{1}}^{0}, S_{B_{2}}^{0}, ..., S_{B_{n}}^{0} )= H^{-1}\Omega > 0 \, \forall i$.

The model system ~\eqref{Eq:3} and ~\eqref{Eq:4} can be put in a compact form as the following

\begin{equation}\label{Eq:Comp}
    \begin{cases}
    	  \frac{dx}{dt}&= \mathcal{A}-\mathcal{B}x\\ 
	  \frac{dy}{dt}&= \mathcal{D}x+\mathcal{E}y
  \end{cases}       
\end{equation}
where $x= [S_M, S_B]^T$, $y= [E_M, I_M, E_{BC}, E_{BSC}, I_{BC}, I_{BSC}, R_{BC}, R_{BSC}]^T$, 

\[
\mathcal{A}=
  \begin{bmatrix}
    \Lambda \\
    \Pi
  \end{bmatrix},
\mathcal{B}=
  \begin{bmatrix}
    diag(\delta+\mu )+M &0 \\
    0&diag(\lambda+\eta+m_B)+P
  \end{bmatrix}
\]
The expressions of $\mathcal{D}$ and $\mathcal{E}$ are included in the Supplementary Information.


\subsubsection{Basic Reproduction Number $R_{0}$ of the patchy model}

 To compute the basic reproduction number, we use the \emph{Next generation method}~\cite{VANDENDRIESSCHE200229}.
Using the notation used in ~\cite{VANDENDRIESSCHE200229}, we can decompose the model system  \eqref{Eq:1} \&  ~\eqref{Eq:2} as $\boldsymbol {\mathcal {F (I)}} - \boldsymbol {\mathcal {V (I)}}$.
$\boldsymbol {\mathcal {F (I)}} \, \, \& \, \, \boldsymbol {\mathcal {V (I)}}$ represent as the flow of new infections and the remaining transfers within and out of the infected classes, respectively.
For the simplicity of our matrix calculation, we have considered the subclinical case as $\gamma_{3_{i}} = 0$ will give the clinical case.
\par
\begin{align*}
\tiny{\boldsymbol {\mathcal {F_{BSC} (I)}}  =
\begin{bmatrix} \frac{c_{1_{1}}I_{BSC_{1}}S_{M_{1}}} {K_{B_{1}}}, & 0, & \frac{\beta_{3_{1}}I_{M_{1}}S_{B{1}}} {K_{B_{1}}}, & \gamma_{3_{1}}R_{BSC_{1}}, & \ldots,&
\frac{c_{1_{n}}I_{BSC_{n}}S_{M_{n}}} {K_{B_{n}}}, & 0, & \frac{\beta_{3_{n}}I_{M_{n}}S_{B{n}}} {K_{B_{n}}}, & \gamma_{3_{n}}R_{BSC_{n}}
\end{bmatrix}}^T
 \end{align*}
Using the notations used before, we have
\begin{align*}
\tiny{\boldsymbol {\mathcal {F_{BSC} (I)}}  = 
\begin{bmatrix} \mu_1S_{M_{1}} , & 0, & \lambda_1S_{B{1}}, & \gamma_{3_{1}}R_{BSC_{1}}, &\ldots,& \mu_n S_{M_{n}}, & 0, & \lambda_{n}S_{B{n}}, & \gamma_{3_{n}}R_{BSC_{n}}
\end{bmatrix}}^T
 \end{align*}
 
\begin{align*}
\tiny{\boldsymbol {\mathcal {V_{BSC} (I)}}}  = 
-\tiny{\begin{bmatrix}
-\gamma_{M}E_{M_{1}}-m_{M_{1}}E_{M_{1}} + \sum m_{1j} E_{M_{j}} -\sum m_{j1} E_{M_{1}}  &\\
\gamma_{M}E_{M_{1}}-m_{M_{1}}I_{M_{1}} + \sum m_{1j} I_{M_{j}} -\sum m_{j1} I_{M_{1}}&\\
-\gamma_{BSC_{1}}E_{BSC_{1}}-m_{B_{1}}E_{BSC_{1}}+ \sum p_{1j} E_{BSC_{j}} -\sum p_{j1} E_{BSC_{1}}&\\
\gamma_{BSC_{1}}E_{BSC_{1}}-m_{B_{1}}I_{BSC_{1}} -\alpha_{3_{1}}I_{BSC_{1}}-d_{BSC_{1}}I_{BSC_{1}} + \sum p_{1j} I_{BSC_{j}} -\sum p_{j1} I_{BSC_{1}}&\\
\vdots&\\
-\gamma_{M}E_{M_{n}}-m_{M_{n}}E_{M_{n}} + \sum m_{nj} E_{M_{j}} -\sum m_{jn} E_{M_{n}}  &\\
\gamma_{M}E_{M_{n}}-m_{M_{n}}I_{M_{n}} + \sum m_{nj} I_{M_{j}} -\sum m_{jn} I_{M_{n}}&\\
-\gamma_{BSC_{n}}E_{BSC_{n}}-m_{B_{n}}E_{BSC_{n}}+ \sum p_{nj} E_{BSC_{j}} -\sum p_{jn} E_{BSC_{n}}&\\
\gamma_{BSC_{n}}E_{BSC_{n}}-m_{B_{n}}I_{BSC_{n}} -\alpha_{3_{n}}I_{BSC_{n}}-d_{BSC_{n}}I_{BSC_{n}} + \sum p_{nj} I_{BSC_{j}} -\sum p_{jn} I_{BSC_{n}}&\\
 \end{bmatrix}}
\end{align*}

Letting $\boldsymbol {F_{BSC}} = [ \partial \boldsymbol {\mathcal {F_{BSC}}}\mid_{(S^{0},\boldsymbol {0} )}]$ and $\boldsymbol {V_{BSC}} = [ \partial  \boldsymbol {\mathcal {V_{BSC}}}\mid_{(S^{0},\boldsymbol {0} )}]$ as the Jacobian matrices evaluated at the disease free equilibrium $(S^{0},\boldsymbol {0} )$.
Following  ~\cite{VANDENDRIESSCHE200229}, the matrix $\mathcal {NGM_{BSC}} = \boldsymbol {F_{BSC}} \boldsymbol {V_{BSC}}^{-1}$ is the next generation matrix for the subclinical birds and it is well defined.

The elements of the Jacobians have the following forms 
\begin{align*}
\frac{\partial \mu_i}{\partial E_{M_{i}}} = \frac{\partial \mu_i}{\partial I_{M_{i}}} = \frac{\partial \mu_i}{\partial E_{BSC_{i}}} =\frac{\partial \lambda_i}{\partial E_{M_{i}}} = \frac{\partial \lambda_i}{\partial E_{BSC_{i}}} = \frac{\partial \lambda_i}{\partial I_{BSC_{i}}} = 0\\
\frac{\partial \mu_i}{\partial E_{M_{j}}} = \frac{\partial \mu_i}{\partial I_{M_{j}}} = \frac{\partial \mu_i}{\partial E_{BSC_{j}}} =\frac{\partial \lambda_i}{\partial E_{M_{j}}} = \frac{\partial \lambda_i}{\partial E_{BSC_{j}}} = \frac{\partial \lambda_i}{\partial I_{BSC_{j}}}  = 0, \, \, \, \,  i \neq j
\end{align*}
\begin{align*}
\frac{\partial R_{BSC_{i}}}{\partial E_{M_{i}}} = \frac{\partial R_{BSC_{i}}}{\partial E_{I_{i}}} = \frac{\partial R_{BSC_{i}}}{\partial E_{BSC_{i}}} =\frac{\partial R_{BSC_{i}}}{\partial I_{BSC_{i}}}=0\\
\frac{\partial R_{BSC_{i}}}{\partial E_{M_{j}}} = \frac{\partial R_{BSC_{i}}}{\partial E_{I_{j}}} = \frac{\partial R_{BSC_{i}}}{\partial E_{BSC_{j}}} =\frac{\partial R_{BSC_{i}}}{\partial I_{BSC_{j}}}=0, \, \, \, \,  i \neq j
\end{align*}
The partial derivatives are evaluated at $(S^{0},\boldsymbol {0})$.
Matrices $\boldsymbol {F_{BSC}}$ and $\boldsymbol {V_{BSC}}$ are $4n \times 4n$ and we express $\boldsymbol {F_{BSC}}$ as $\boldsymbol {F_{BSC}} = diag[\boldsymbol {F_{BSC_{ii}}}]$,
with $i = 1, 2, \ldots, n$ and 
\begin{align*}
\boldsymbol {F_{BSC_{ii}}} = 
\begin{bmatrix}
0 & 0 & 0 & 0 \\
0 & 0 & \frac{\partial \lambda_i}{\partial I_{M_{i}}}S_{B}^{0} & 0 \\
0 & 0 & 0 & 0 \\
\frac{\partial \mu_i}{\partial I_{BSC_{i}}}S_{M}^{0}  & 0 & 0 & 0 
 \end{bmatrix}
\end{align*}
and 
$\boldsymbol {V_{BSC}} = [\boldsymbol {V_{BSC_{ij}}}]$, where 
\begin{align*}
\boldsymbol {V_{BSC_{ij}}} = diag
\begin{bmatrix}
-m_{ij} & -m_{ij} & -p_{ij} & -p_{ij}
 \end{bmatrix}
 ;  \, \, \, \,  i \neq j
\end{align*}
and 
\begin{align*}
\boldsymbol {V_{BSC_{ii}}} =
\begin{bmatrix}
\gamma_M+m_{M_{i}}+ \sum m_{ji} & -\gamma_M & 0 & 0 \\
0 & m_{M_{i}}+ \sum m_{ij} & 0 & 0 \\
0 & 0 & \gamma_{BSC_{i}}+m_{B_{i}}+ \sum p_{ji} &  -\gamma_{BSC_{i}}\\
0 & 0 & 0 & m_{B_{i}}+\alpha_{3_{i}}+d_{BSC_{i}}+ \sum p_{ji}
\end{bmatrix}
\end{align*}

Similarly the next generation matrix ($\mathcal {NGM_{BC}}$) associated with the clinical birds is $\boldsymbol {F_{BC}} \boldsymbol {V_{BC}}^{-1}$.
After including clinical and subclinical birds in different patches, the next generation matrix ($\mathcal {NGM}$) of the system  ~\ref{Eq:1} \&  ~\ref{Eq:2} is following
\begin{align*}
\mathcal {NGM} = \left[
    \begin{array}{c:c}
        \mathcal {NGM_{BC}} &
        \mathcal {NGM_{BSC}}
    \end{array}
\right]
\end{align*}

So, the basic reproduction number is  
\begin{equation} \label{R0Met}
\mathcal {R}_{0} = \rho(\mathcal {NGM})
\end{equation}
where $\rho$ is the spectral radius of the matrix $\mathcal {NGM}$.
According to  ~\cite{VANDENDRIESSCHE200229}, the local stability of the disease free equilibrium $E_0 = (S^{0},\boldsymbol {0})$ is governed by $\mathcal {R}_{0}$.
If $\mathcal {R}_{0} < 1$, then $E_0$ is asymptotically unstable and unstable whenever $\mathcal {R}_{0} > 1$.


\section{Simulation Results}
\subsection{Impact of Mosquito Mobility}
Most of the literature in the area of vector-borne disease modelling centres around the movement on the long distance and long duration travelling of the host species only.
In those models, we can overlook the vector mobility but in our current effort we include the short and small scale flights of the vectors of WNV.
The influence of the vectorial capacity and the movements of the mosquitoes are important features to the potential spread and hence sustaining WNV  \cite{MOULAY2013129, Ciota}.
Therefore, we conceive that the daily mobility and the short journeys carried by the vector species can not be ignored.
It is of our interest to acknowledge that smaller temporal scale of WNV transmission can potentially include the new aspects in the spread of WNV in Germany.

\begin{figure}[H]
\centering
\subfloat[]{
  \includegraphics[width=0.5\textwidth]{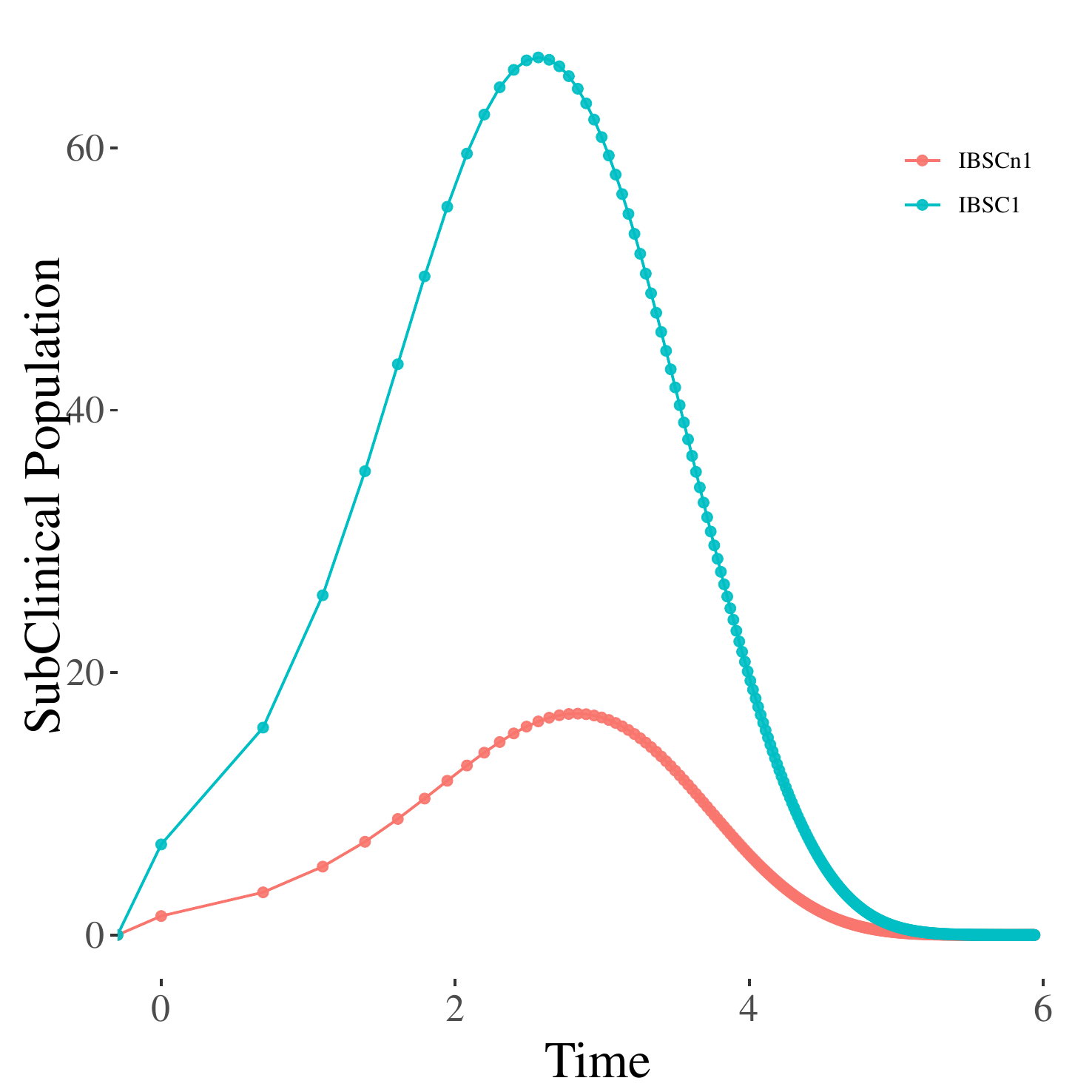}
   \label{Fig:IBSCvsMos1}
}
\subfloat[]{
  \includegraphics[width=0.5\textwidth]{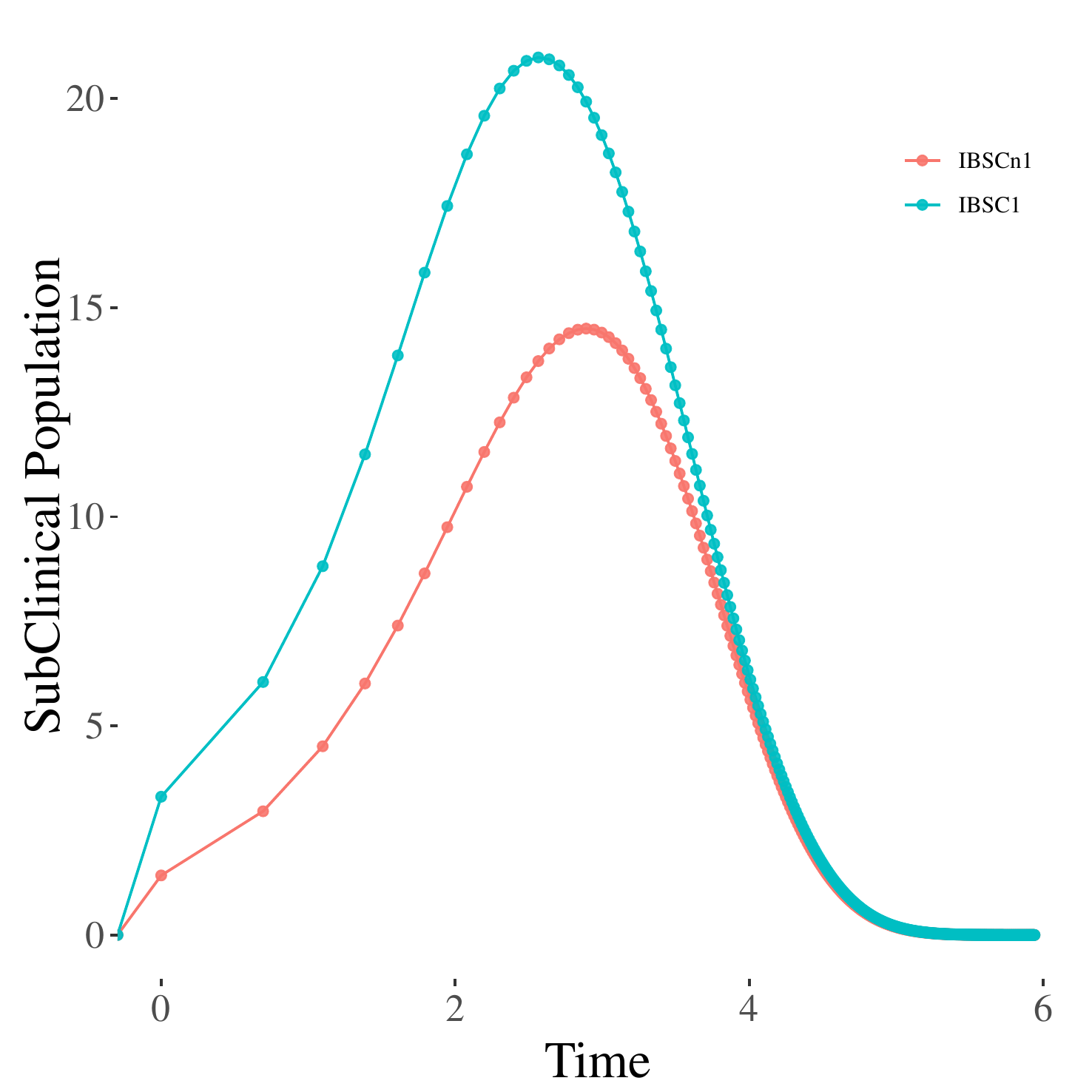}
   \label{Fig:IBSCvsMos2}
} 
\caption{
Graphical representations of effect of mosquito mobility in the time evolution of infected subclinical bird population.
(a) In this simulation we have chosen the range of mosquito mobility as the average maximum distance.
(b) In this simulation we have chosen the range of mosquito mobility as the minimum of maximum distance as described in the section \ref{Vector Mobility Network}. 
}
\label{Fig:MosMoveImp}
\end{figure}
The expression of $\mathcal {R}_{0}$ for the metapopulation model is too long and complicated enough to perform theoretical analysis to understand the influence of the movements
of the mosquitoes.
So, we decide to consider only the subclinical birds and examine the influence of the mosquito movements.
Here, we simulate the situation when an infected local subclinical bird is introduced to a completely susceptible population with the different flight range movements of the mosquitoes.
We consider two cases for the simulations, one with the inclusion of local mosquito mobility and another one without the mosquito migration. 
In both cases, we keep the subclinical birds movement enabled.
The importance of local mosquito interactions is visible in the Figure:~\ref{Fig:MosMoveImp}. 
It shows the number of subclinical infected birds versus time.
$I_{BSC1}$ stands for the infected subclinical birds in the patch number 1 and  $I_{BSCn1}$ stands for the subclinical birds in the patch number 1 but without mosquito movements.
It is interesting to observe that in the scenario when we do not include the mosquito movements, the infection spreading process takes a bit longer time and the peak is relatively flat whereas when we include the mosquito mobility, the infection spreads quicker than previously considered case and the peak is sharp and concentrated with subclinical birds. 
For the experimental purpose, we change the maximum flight ranges of mosquitoes and the influence of the mosquito movements are clearly visible in the 
Figure: \ref{Fig:IBSCvsMos1} and Figure: \ref{Fig:IBSCvsMos2}. 
Higher range of mosquito movements facilitate the potential transmission of WNV in the local birds population.

\subsection{Impact of Bird Movements}
The shape of the epidemiology of WNV spread in Germany is potentially governed by the movements of the  interacting species especially the birds as a prime host.
It is widely accepted that the movements of the birds likely to facilitate the dissemination of WNV in Germany \cite{Ute, ZIEGLER201939}.
In this section we explore the influence of the birds (both local and long dispersal birds) movements on the potential spatial spread of WNV in Germany.
With the intention to comprehend the impact of bird movements, we set up some simulations under certain assumptions.
In the following sections we elaborate them.
\subsubsection{Impact of Local Bird Movement}\label{ImpactLocBird}
In this section we explore the influence of the local bird movements on the basic reproduction number of the patchy model system  \eqref{Eq:1} \&  ~\eqref{Eq:2}.
To keep our findings simple, we just consider two-patch model consisting only clinical birds and the corresponding basic reproduction number $\mathcal {R}_{0}$ given in  \eqref{R0Met}.
The analytical and the symbolic computations related to the clinical and subclinical birds are the same.
Given the complex and the long expression of $\mathcal {R}_{0}$ (Included in the Supplementary Information) even for the two-patch model, it is rather difficult to quantify theoretically  the impact of the local bird migration in $\mathcal {R}_{0}$ and in the dissemination of WNV from one habitat patch to another.
So, we take the refuge of mathematical simulations under different conditions to understand it after following \cite{RUAN201765}.

\begin{figure}[H]
\centering
\subfloat[]{
  \includegraphics[width=0.5\textwidth]{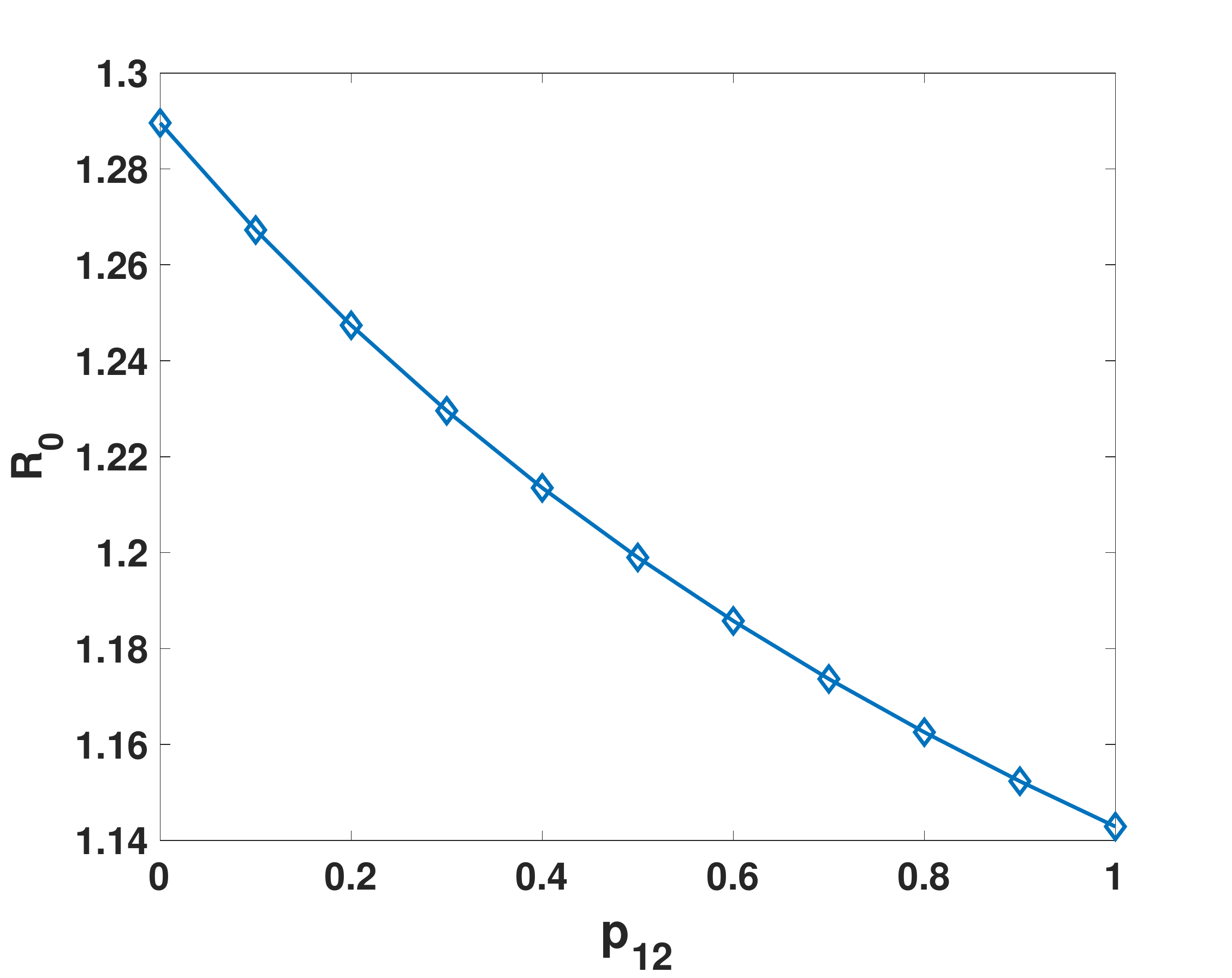}
   \label{Fig:p_12vsR_0}
}
\subfloat[]{
  \includegraphics[width=0.5\textwidth]{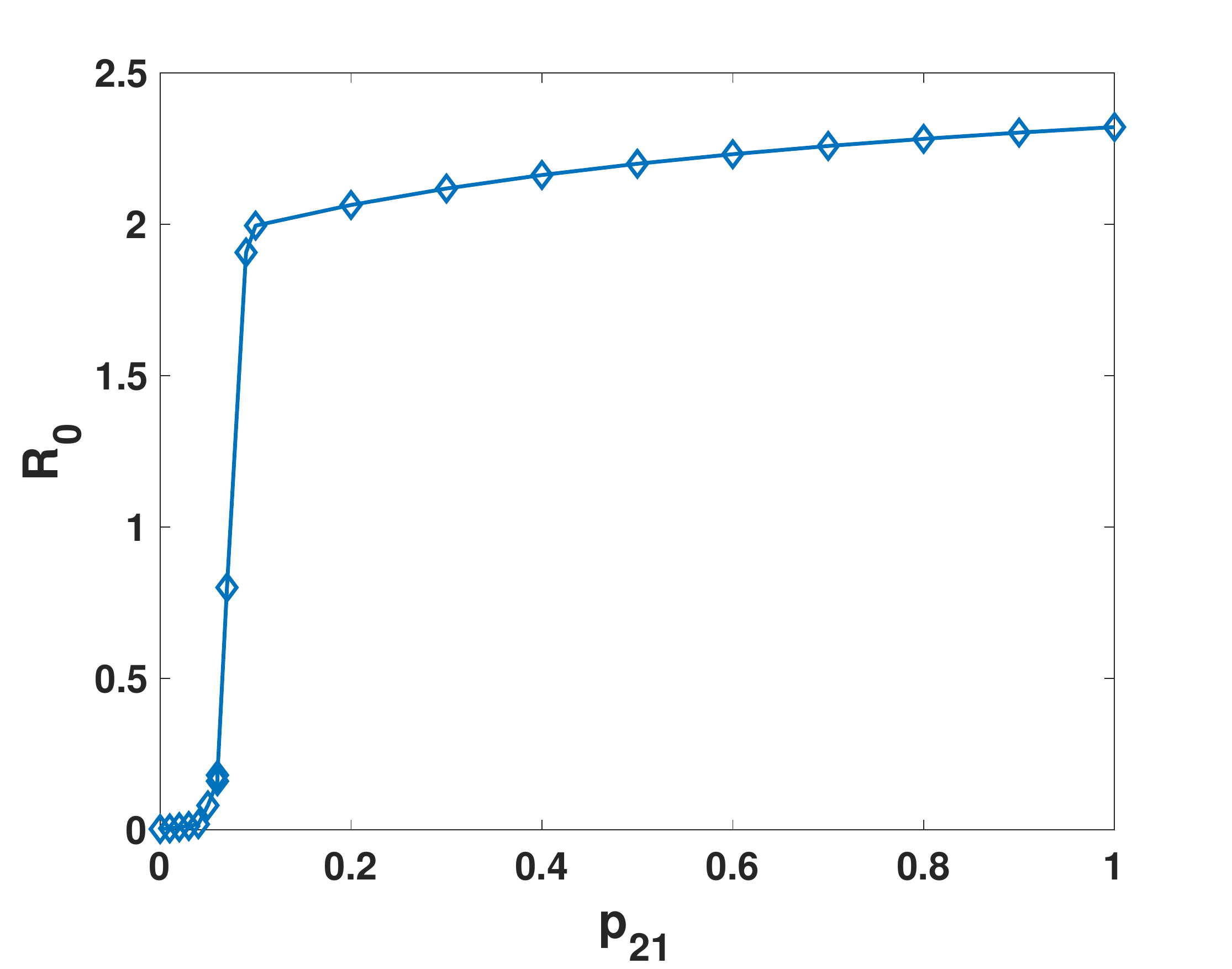}
   \label{Fig:p_21vsR_0}
} 
\caption{
Graphical representations of local bird movements on $\mathcal {R}_{0}$ for a two patch model.
The symbolic computations have been performed in \cite{dCode}.
(a) Impact of local bird (subclinical) migration from the second patch to the first patch and 
(b) Impact of local bird  (subclinical) migration from the first patch to the second patch.
}
\label{Fig:PijvsR0}
\end{figure}
To test the influence on local bird movements, we run the model for two patches using the following parameters:  $m_{B_{1}} = m_{B_{2}} = 0.016$, $\gamma_{BSC_{1}} = \gamma_{BSC_{2}} = 0.567$, $\alpha_{3_{1}} = \alpha_{3_{2}} = 0.182$, $K_{M_{1}} = K_{M_{2}} = 1000$, 
$K_{B_{1}} = K_{B_{2}} = 100$,  $c_{1_{1}} = c_{1_{2}} = 0.18$, $d_{IBSC_{1}} = d_{IBSC_{2}} = 0.5$,  $m_{M_{1}} = m_{M_{2}} = 0.4141$. 
Only the transmission parameters are different for the subclinical birds. 
For the second patch it is 0.88 and for the first patch it is 0.78, mosquito migration rates are kept the same for both the patches as 0.9.
We assume that the basic reproduction number ($\mathcal {R}_{0_{1}}$)of patch $1$ is less than $1$ and for the patch $2$ it is ($\mathcal {R}_{0_{2}}$) greater than $1$ i.e. $\mathcal {R}_{0_{1}} = 0.87465$ and $\mathcal {R}_{0_{2}} = 1.090$.
Here, we carry out the simulations to investigate the impact of the local bird migrations on the basic reproduction number. 
For this purpose, we keep all the parameters same except for the transmission parameters from the mosquitoes to the clinical birds for the two-patch model.
From the magnitude of the basic reproduction number it is clear that WNV is endemic in the patch $2$ and in the patch $1$, it will die out.
First we let, $p_{21} = 0.91$ and let let $p_{12}$ to vary. 
Then we keep $p_{12}= 0.9$ and vary $p_{21}$ to observe the influence of the clinical birds on the magnitude of the basic reproduction number.
From the Figure: \ref{Fig:p_12vsR_0}, it is evident that with the increase of migration from the patch $2$ to the patch $1$, the basic reproduction number reduces whereas we can witness the opposite of this phenomenon in the Figure: \ref{Fig:p_21vsR_0}, where with the increase of $p_{21}$ yields the increase in the basic reproduction number of the two patch model.\par
These simulations possibly give us the glimpse of the complexity of the potential spread of WNV from one endemic patch to another one.
The immigration of the birds can trigger a series of such expansion of WNV and this kind of complexities can not be explained through a localised model~\cite{BHOWMICK2020110117}.

\subsubsection{Impact of Long Distance Bird Movement}
The authors in \cite{liu2006modeling} investigate that the dispersal of long range birds actually establish the spatial spread of WNV.
It is plausible that the  acquisition of WNV can happen during dispersal of long range birds  \cite{Reed}.
In this simulation exercise, we include the movements of the local birds and mosquitoes but discard the movements of the long range dispersal birds to understand the importance of 
the movements of long range dispersal birds in introducing and sustaining the bite of WNV in the new places.

\begin{figure}[H]
\centering
\subfloat[Subfigure 1 list of figures text][]{
\includegraphics[width=0.5\textwidth]{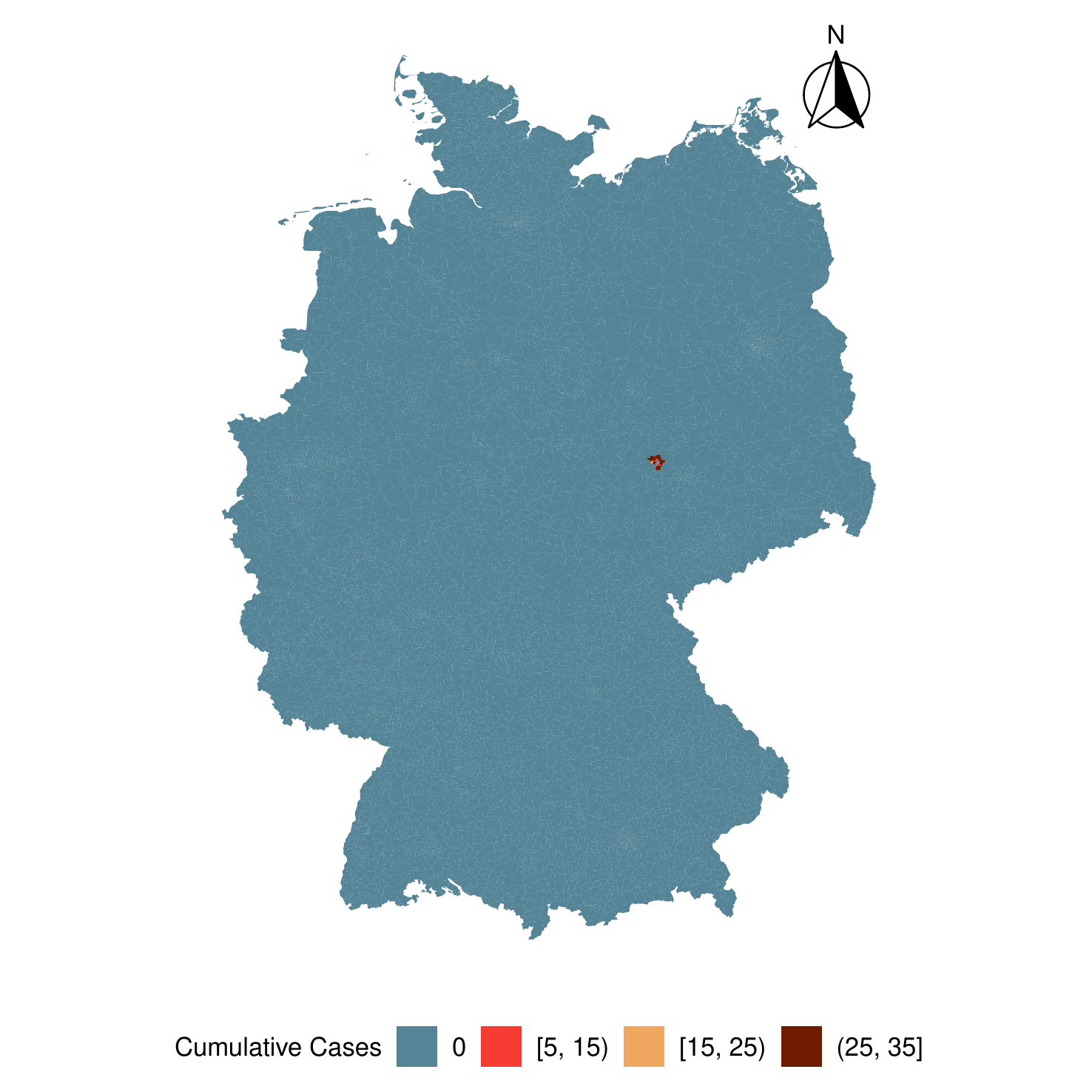}
    \label{Fig:LDD1}
}
\subfloat[Subfigure 2 list of figures text][]{
\includegraphics[width=0.5\textwidth]{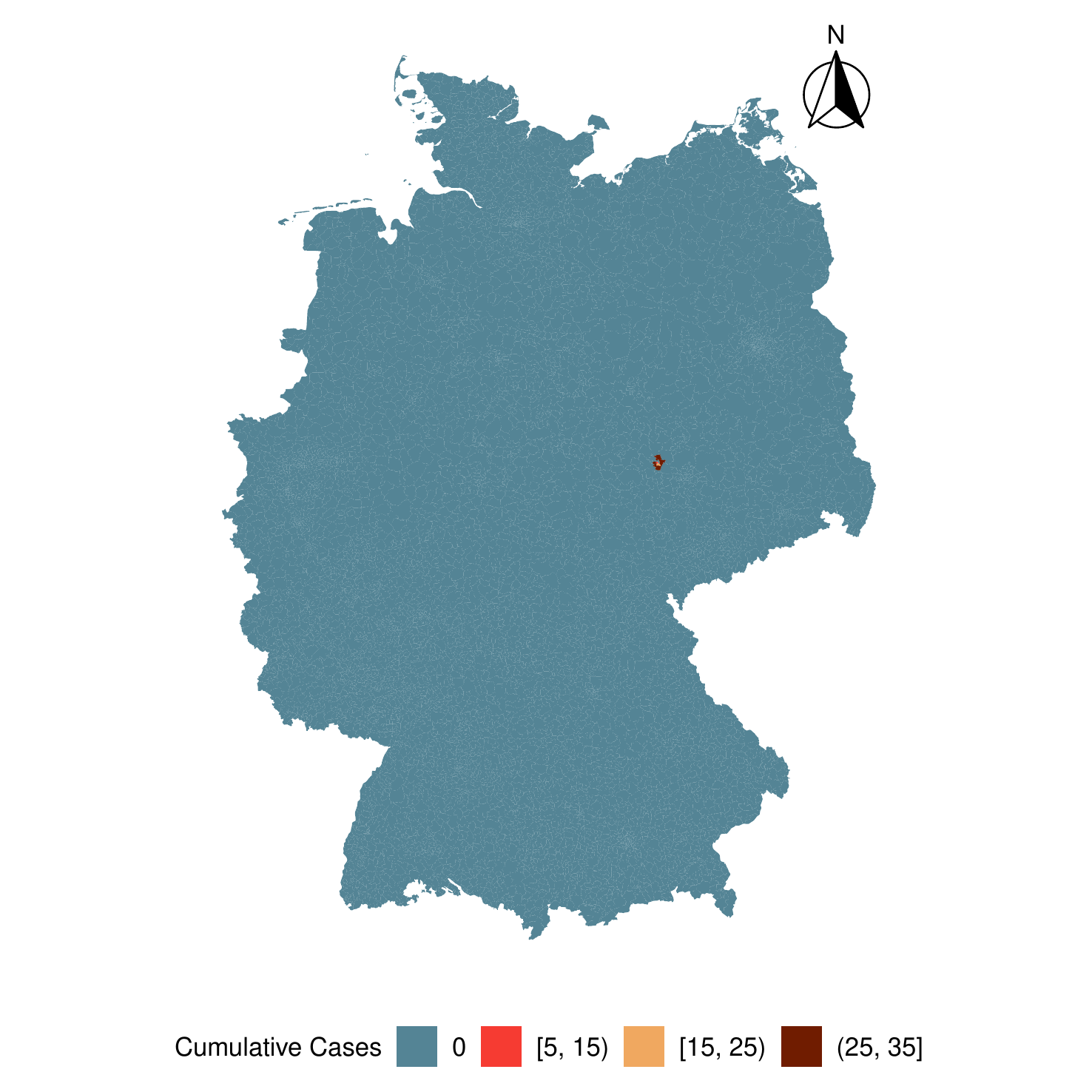}
 \label{Fig:LDD2}
}
\qquad
\subfloat[Subfigure 2 list of figures text][]{
\includegraphics[width=0.5\textwidth]{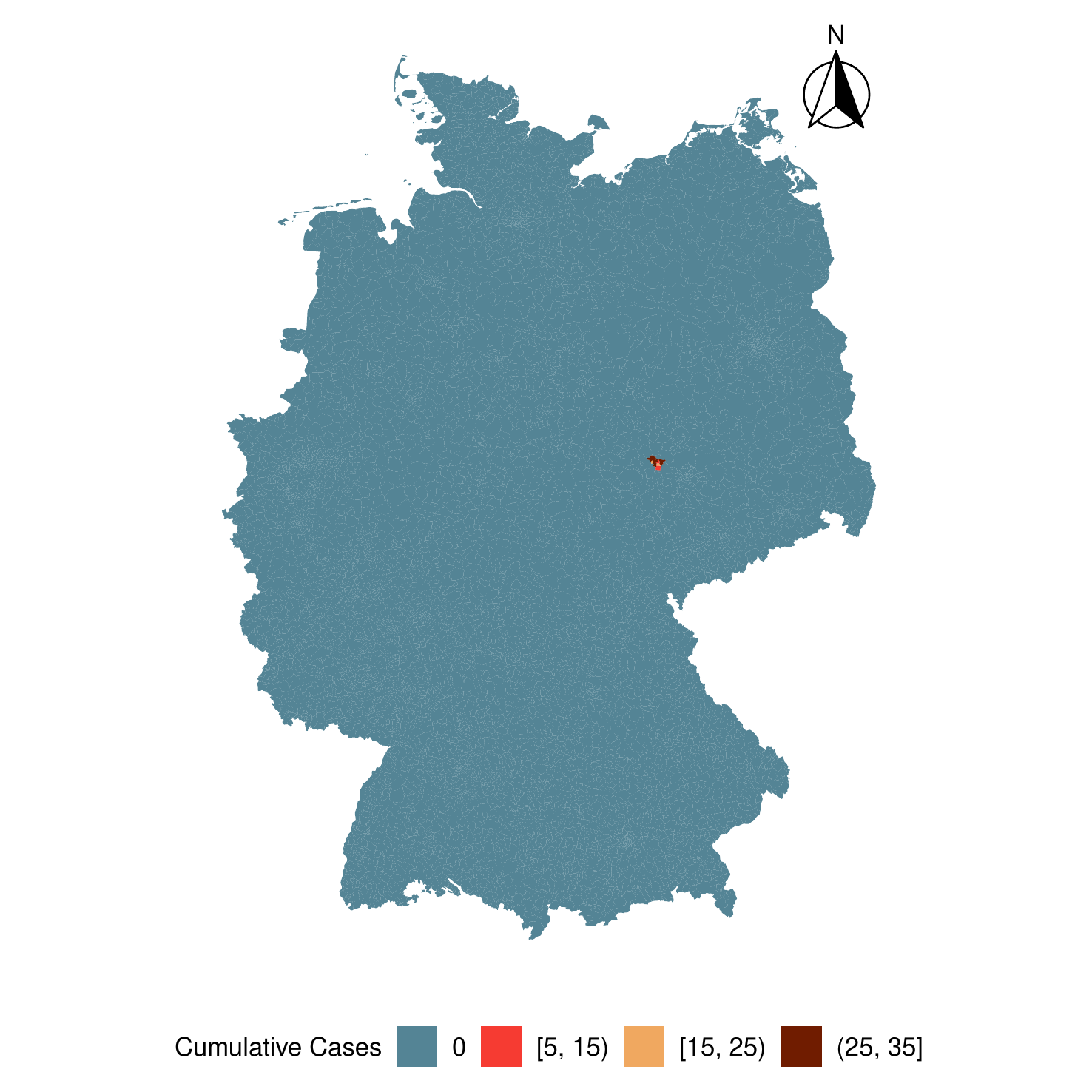}
 \label{Fig:LDD3}
}
\subfloat[Subfigure 2 list of figures text][]{
\includegraphics[width=0.5\textwidth]{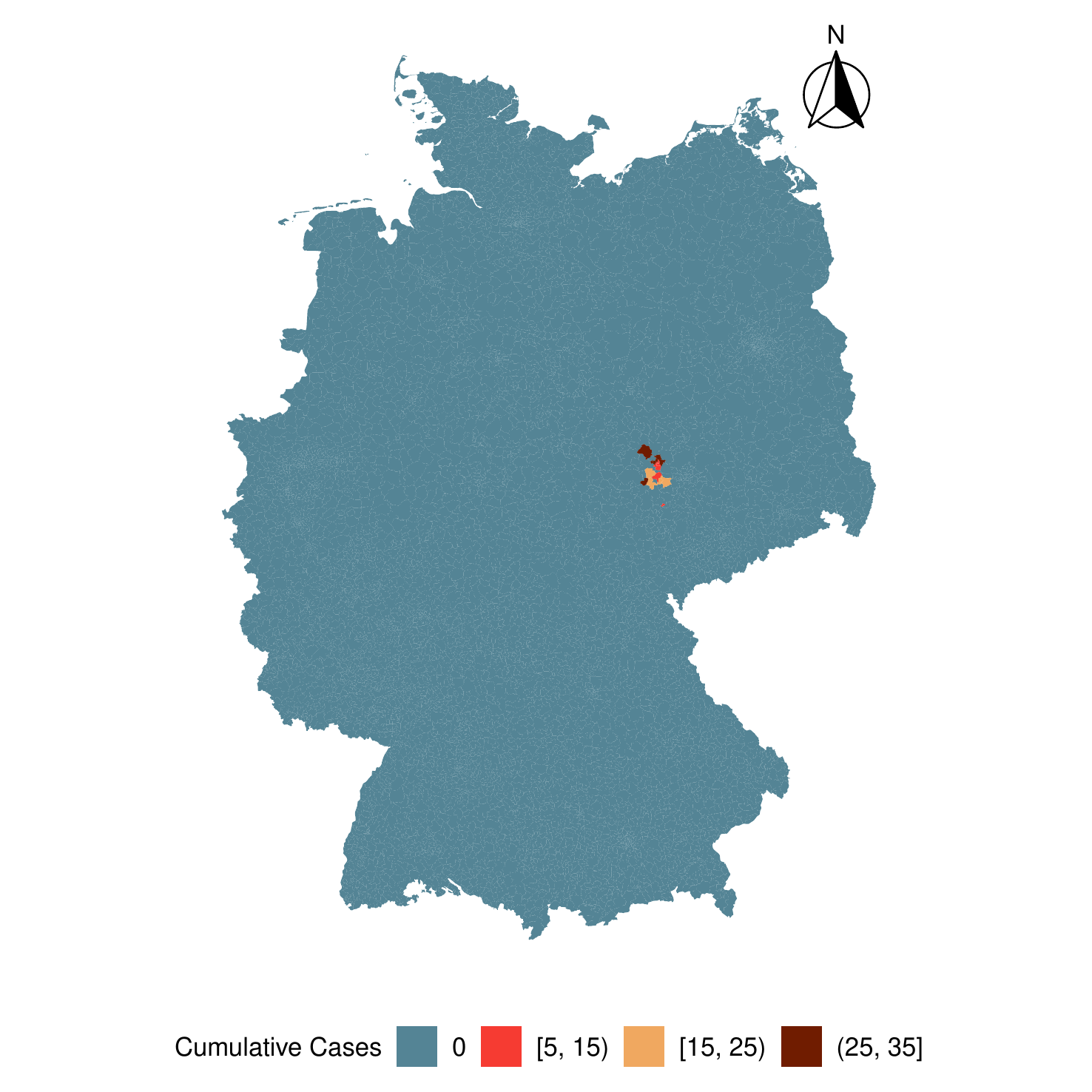}
 \label{Fig:LDD4}
}
\caption{(a) Choropleth maps of  potential spatial spread of WNV in Germany after season 1 without the movements of long range dispersal birds with $0.75$ and $0.001$ as shape and scale parameters
(b) Choropleth maps of  potential spatial spread of WNV in Germany after season 1 without the movements of long range dispersal birds with $0.95$ and $0.01$ as shape and scale parameters
(c) Choropleth maps of  potential spatial spread of WNV in Germany after season 1 without the movements of long range dispersal birds with $1.0$ and $0.1$ as shape and scale parameters
(d) Choropleth maps of  potential spatial spread of WNV in Germany after season 1 without the movements of long range dispersal birds with $2.8$ and $1.0$ as shape and scale parameters
 }
\label{Fig:LongRange}
\end{figure}
We introduce $I_{\mathrm{Inf}}(= 1)$ number of infected long range dispersal bird in selected city of Halle as the first case of WNV was reported in that city \cite{ZIEGLER201939}.
We employ different dispersal kernels of the local birds to spatially map the potential spread of WNV under the above mentioned assumptions to run the simulations with the different parameter values as described in Section \ref{LocalBird}.
It is noticeable that without the inclusion of movements of the long range dispersal birds, the spread of WNV  is rather concentrated and the pace at which transmits is comparatively slow.
From this simulation experiment we can infer the potent role of long range dispersal birds to introduce WNV into the new territories.
We have included the graphical representations of Weibull distribution parameters in Supplementary Information.
We compute structural similarity (SSIM) index \cite{Massaro} score as mentioned in the Section \ref{SpatialSpread} for each epidemiological simulation and append the findings in the Supplementary Information.

\section{Vector Control }
One of the main applications of models for the potential spread of WNV among the birds is to design of possible control strategies and create buffer zones to prevent, or  minimise the spread of WNV.
For the demonstration purpose we consider only the movement between two patches of the clinical birds. 
Mosquito movement is also considered.
We follow~\cite{Lewis2019} to assess mathematically the feasible control strategies.
Consider the model system~\eqref{Eq:1} and~\eqref{Eq:2} and we take account of the clinical birds and their movements between two patches only.
The  next generation matrix for the clinical birds turns out to be in the following form 
\begin{align*}
\mathcal {NGM_{BC}} =
\begin{bmatrix}
0 & 0 & a_1 & a_2 \\
0 & 0 & a_3 & a_4 \\
a_5 & a_6 & 0 &  0\\
a_7 & a_8 & 0 & 0
\end{bmatrix}
\end{align*}
The explicit mathematical forms of $a_i$, $i= 1: 8$ are included in the Supplementary Information. 

$\mathcal {NGM_{BC}}$ can be readily put into the block matrix form as following:
\begin{align*}
\mathcal {NGM_{BC}} =
\begin{bmatrix}
\mathbf {0} & \mathbf{A_1} \\
\mathbf{A_2} & \mathbf {0} \\
\end{bmatrix}
\end{align*}
where
\[
\mathbf{A_1} = 
\begin{bmatrix}
a_1 & a_2 \\
a_3 & a_4 \\
\end{bmatrix} \,
\mathbf{A_2} = 
\begin{bmatrix}
a_5 & a_6 \\
a_7 & a_8 \\
\end{bmatrix}
  \]
Then according to  \cite[Theorem 8 (1)]{Lewis2019}, targeting either the vector or the host population to control WNV should be the effective strategy while counting on the cost of corresponding one-group target strategies.
According to Theorem 8, it is better to apply all available resources to target only one population only.
In order to lessen the burden of WNV,  the mosquito population needs to be reduced below a certain threshold. 
In our case mosquito control is only option to explore.
According to  \cite{Lewis2019}, the type reproduction number ($\mathcal{T}_M$) to control the mosquito population is given by: 
$ \mathcal{T}_M = \frac{\mathbf{A_1}\mathbf{A_2}}{1-p_1}$, where $p_1$ is the fraction of the vector population to be controlled. 
Therefore, more than the $1- \frac{1}{\mathcal{T}_M}$ fraction of mosquitoes should be targeted to eradicate the impact of WNV in case of two patch populations.
Using the parameters described in \ref{ImpactLocBird}, more than $56.34\%$  of the mosquito population should be reduced to potentially stop the spatial transmission of WNV in the 
two patch population model.

\section{Spatial spread in Germany}\label{SpatialSpread}
The study area comprised of 11,054 German Gemeinden (Municipalities).
We use the deterministic metapopulation~\ref{Eq:3} \& ~\ref{Eq:4} and the equations associated with the movement of birds (included in the Supplementary Information) to apprehend the WNV transmission in the local bird populations in each \textit{Gemeinde}.
We are caught increasingly between the complexity of the simulations and the possibility to decipher the potential spatial spread of WNV in Germany qualitatively. 
Birds are grouped according to their health status as depicted in \ref{Fig:Metapop1}.
We explicitly represent the vector population and the mobility.
Unfortunately, we do not have the population distribution of birds across the different Gemeinden level so we have assumed uniform number of local birds as in \cite{BHOWMICK2020110117}.\par 
At first, we keep all the local birds initially susceptible in all the Gemeinden.
We introduce $I_{\mathrm{Inf}}(= 1)$ number of infected long range dispersal bird in selected Gemeinden.
For the first season simulation the infection is seeded in the city Halle as the first case of WNV was reported in that \cite{ZIEGLER201939}.\par
Between-Gemeinden movements of vectors and the hosts take place on three different contact networks: (i) the vector network, representing the mosquito flies
(ii) the local birds network, represents the movements of the local birds in a habitat patch and (iii) the long distance movement birds network, represents sporadic movements of birds ranging distances between 0 and 500 km pathways.

\begin{figure}[h!]
    \centering
     \includegraphics[width=0.5\textwidth]{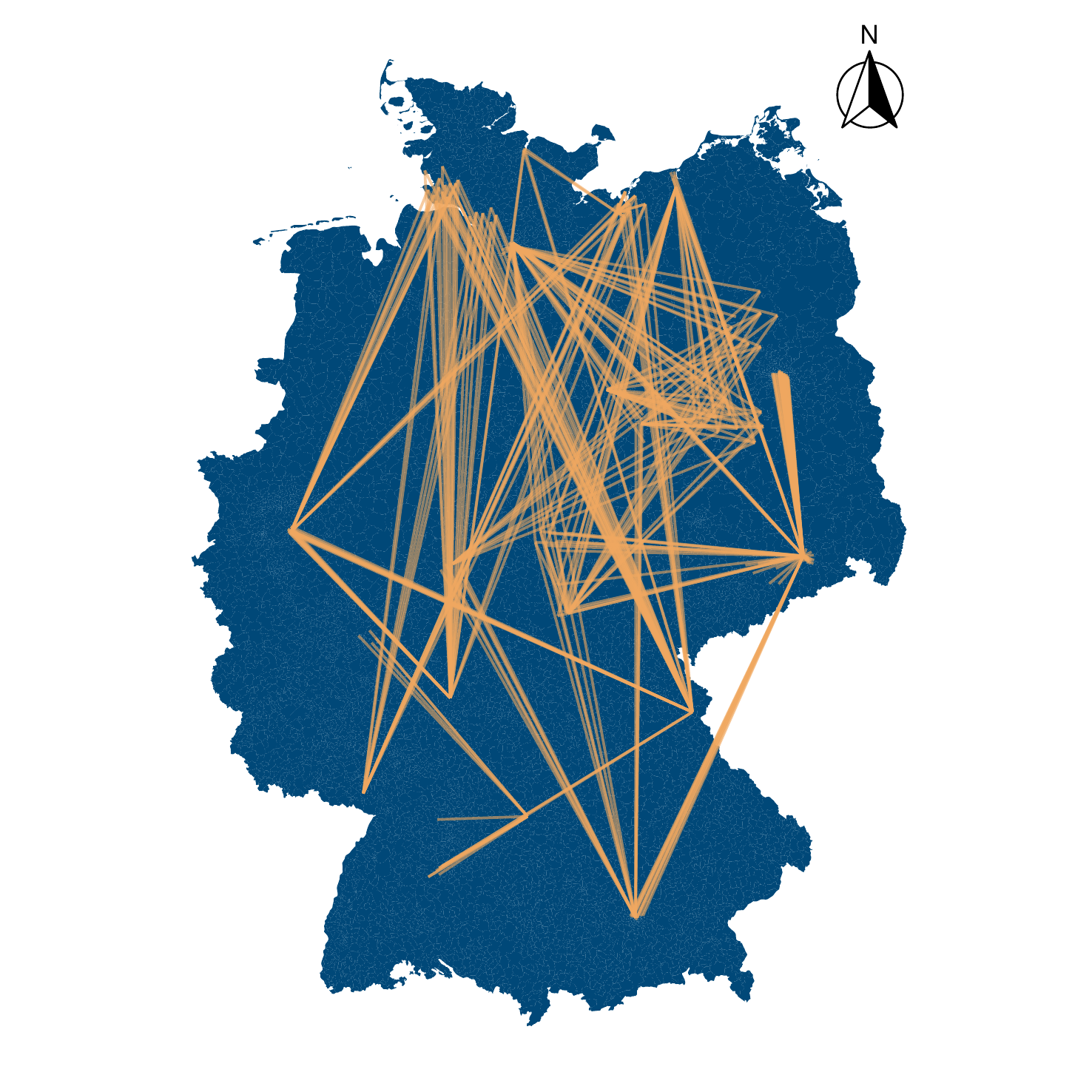}
    \caption{Dispersal networks for power-law distance based kernel of long range dispersal birds in Germany according to the Algorithm: \ref{Alg:MigBirdNet}}
    \label{Fig:MignetGer}
\end{figure}

The nodes are the centroid of Gemeinden and the links amongst them are formed after employing Algorithm: \ref{Alg:MosNet}, Algorithm: \ref{Alg:BirdNet} and Algorithm: \ref{Alg:MigBirdNet} for the three
distinct networks.
In Figure \ref{Fig:MignetGer}, we depict the long range dispersal birds pathway being constructed through the power-law distance based kernel.
\par
In the simulation, we take note of the source of infection of each newly infected Gemeinden. 
Gemeinden that are previously free of WNV might be infected through different networks of concern.
We then take account of cumulative number of infected birds, both the local and the long range dispersal birds in each season of WNV circulation per Gemeinden in Germany.
\begin{figure}[H]
\centering
\subfloat[Subfigure 1 list of figures text][]{
\includegraphics[width=0.5\textwidth]{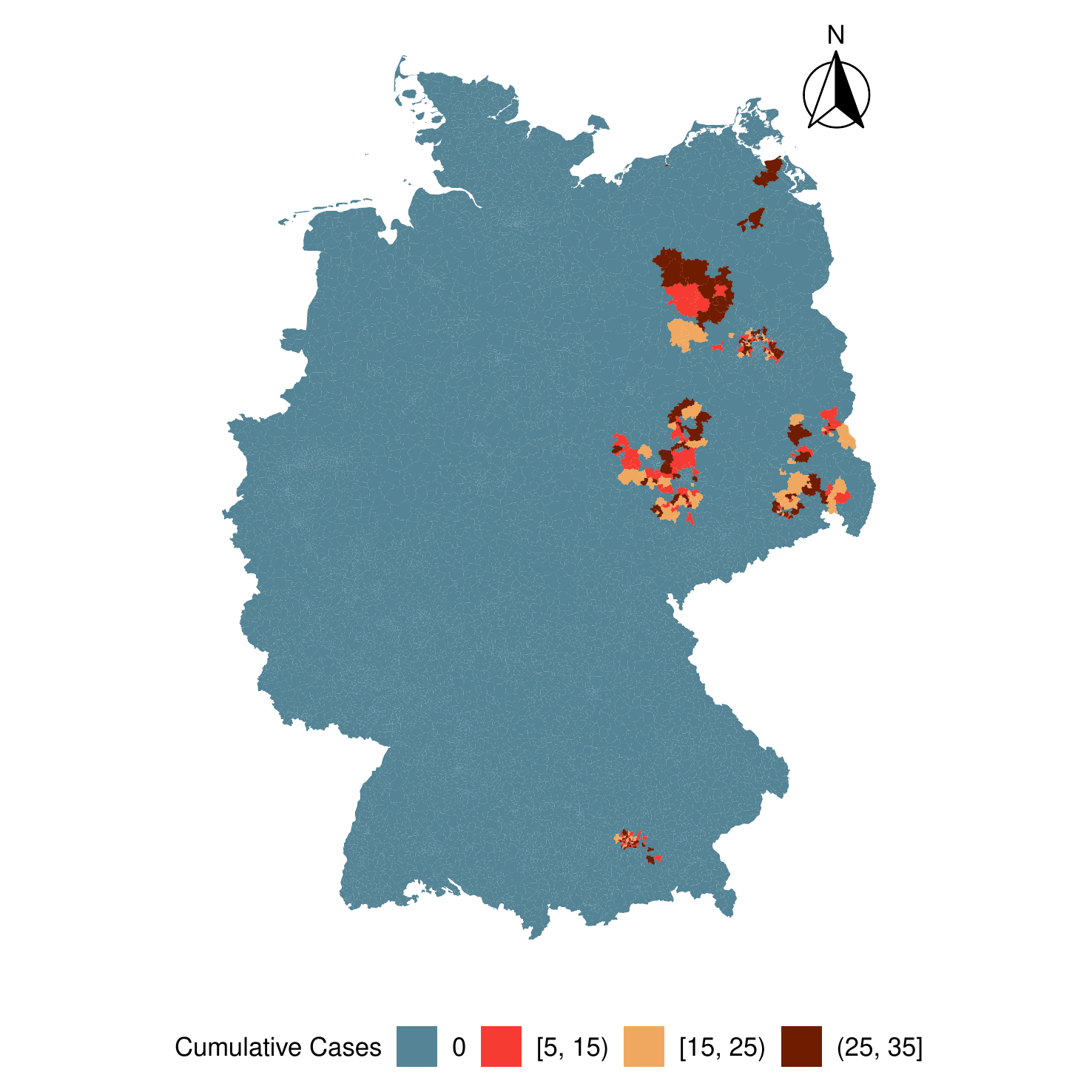}
    \label{Fig:Spatial1}
}
\subfloat[Subfigure 2 list of figures text][]{
\includegraphics[width=0.5\textwidth]{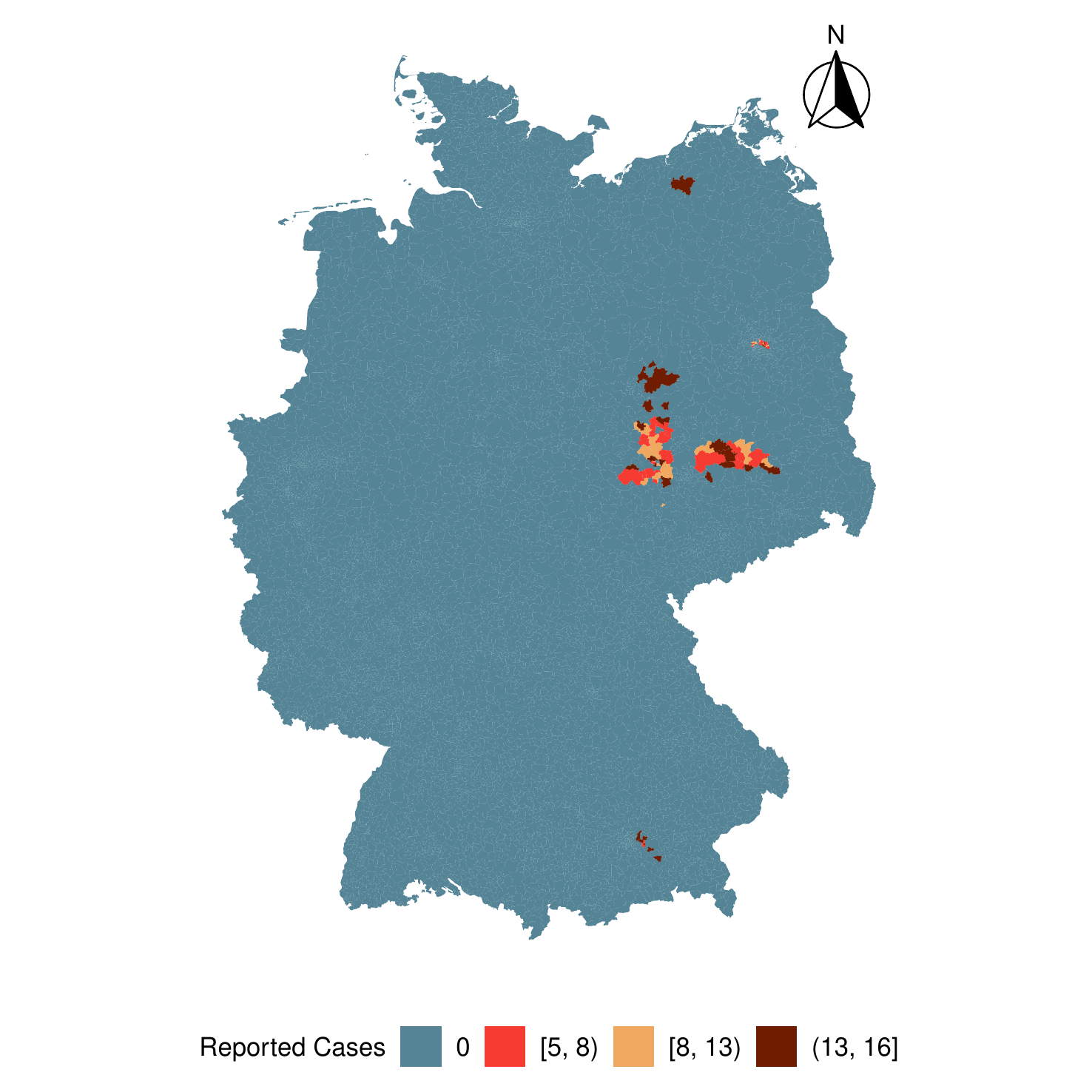}
 \label{Fig:Spatial2}
}
\qquad
\subfloat[Subfigure 2 list of figures text][]{
\includegraphics[width=0.5\textwidth]{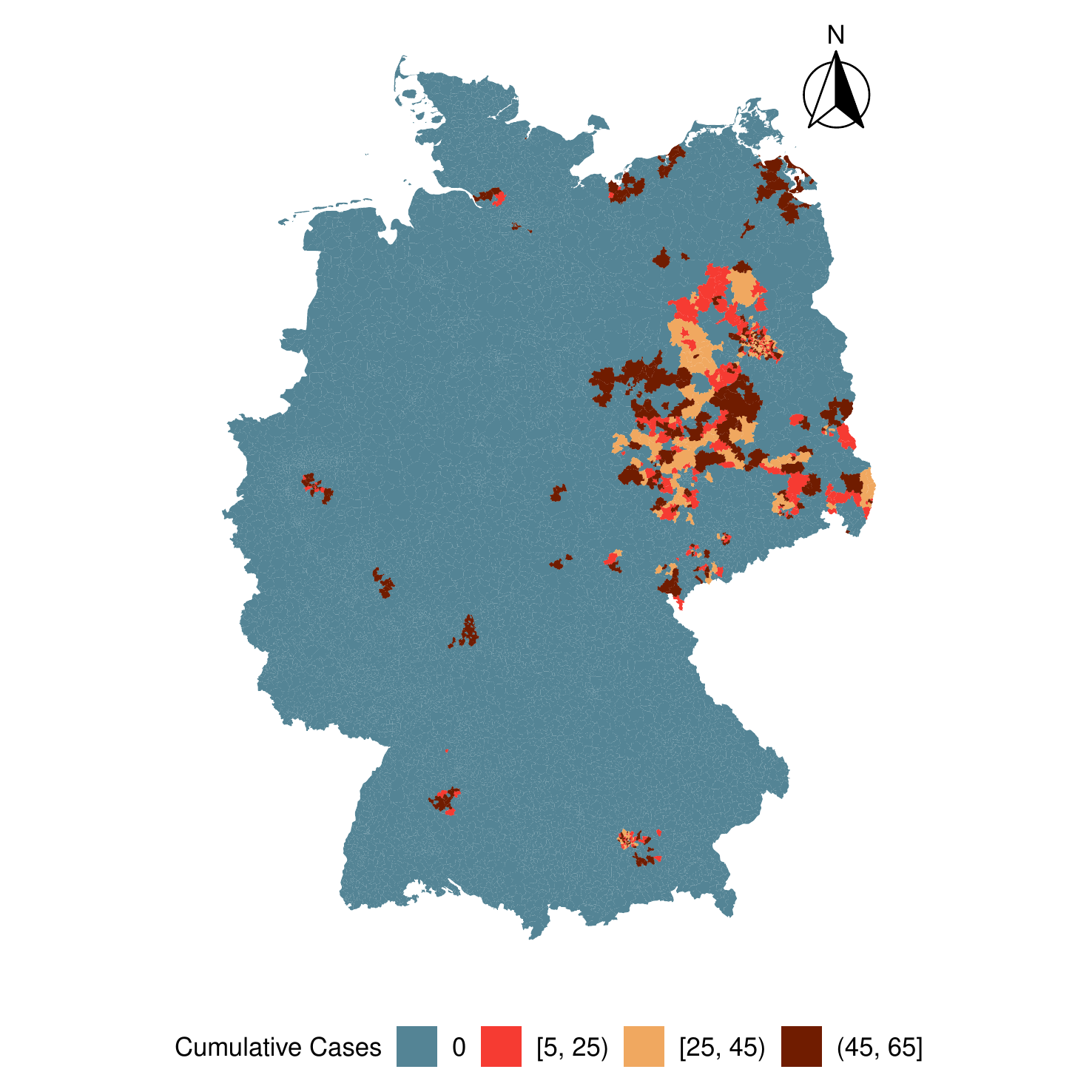}
 \label{Fig:Spatial3}
}
\subfloat[Subfigure 2 list of figures text][]{
\includegraphics[width=0.5\textwidth]{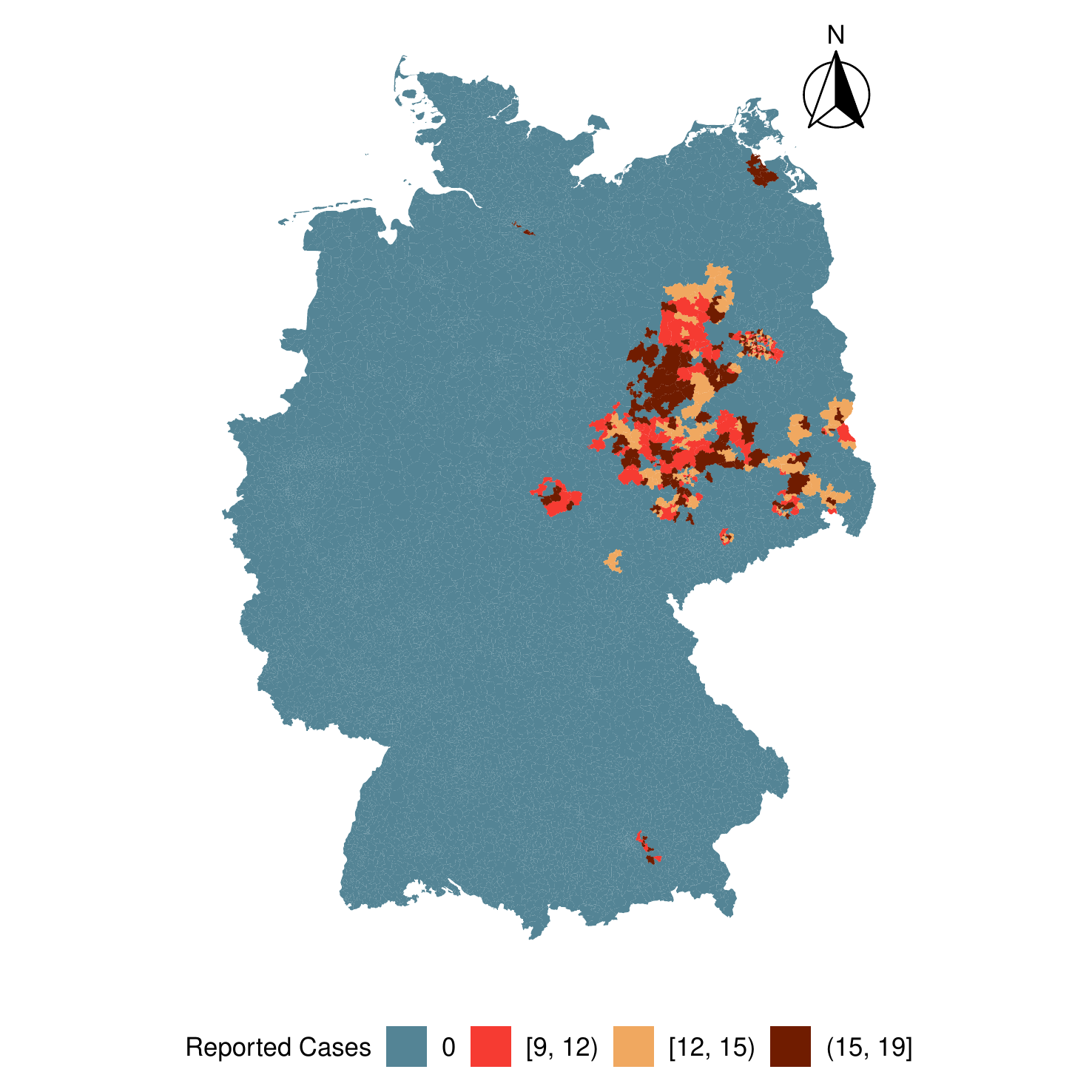}
 \label{Fig:Spatial4}
}
\caption{(a) Choropleth maps of  potential spatial spread of WNV in Germany after season 1 
(b) Choropleth maps of cumulative reported cases of WNV in Gemeinden, Germany in 2018, data from \cite{TierSeuchenInformationsSystem}
(c) Choropleth maps of  potential spatial spread of WNV in Germany after season 2
(d) Choropleth maps of  cumulative reported cases of WNV in Gemeinden, Germany in 2019, data from \cite{TierSeuchenInformationsSystem}
 }
\label{Fig:Spatial}
\end{figure}
Simulated infection spread data can allow us for the spatial projection on the map of Germany.
Complementarily to Figure: \ref{Fig:MignetGer}, we present in Figure: \ref{Fig:Spatial} further results on the temporal and spatial evolution of the disease dynamics.
Here, we compute, through simulations of the model, the cumulative  number of infected birds within each region assuming that the disease propagates from the seeding zone to the other places through the combinations of the different spatial dispersal networks.
In the Figure: \ref{Fig:Spatial1} and Figure: \ref{Fig:Spatial2}, we show the spatial distribution of potential WNV cases while including the probable spatial spread of WNV in Germany and the cumulative reported cases of WNV in the year $2018$ according to the counting per  Gemeinden.
We perform the same for the next season.
By the end of the WNV season 1, apparently the spatial spread of WNV infection is mostly confined within the Eastern zone of Germany with some spontaneous cases in the north and in the southern zones. \par
Figure: \ref{Fig:Spatial3} represents the simulated possible spatial spread of WNV in Germany and in Figure: \ref{Fig:Spatial4}, cumulative reported cases of WNV in the year $2019$ according to the counting per  Gemeinden are presented.

Compared to the observed outbreak regions, the affected regions in the outbreak model are very similar in 2018 (figure \ref{Fig:Spatial1} and figure \ref{Fig:Spatial2}). 
Nevertheless, the outbreak region in the model is bigger than in the observed data.
In 2019, the model shows several long distance spread, that hasn't been observed (compare figure \ref{Fig:Spatial3} and figure \ref{Fig:Spatial4}). 
The spatial spread align with the theoretical predictions and possibly highlight the close relationship between the spatial spread of WNV into the newer zones and the flyways of long range dispersal birds, even though the newly infected zones are relatively far geographically.
Finally, it is worth mentioning that there are two major differences between our model and the realty.
First, the observed data rely on dead birds found in the environment or in husbandry and therefore there might be a reporting bias what is not known precisely. 
The model predictions shows the total number of infected birds.
Secondly, the direction of the spatial spread in our model is not weighted, in contrast to the reality where birds fly to the feeding places. 
Additionally, we would like to emphasise that many unknown factors, e.g. long distance flyway directions, could not be taken into consideration.
However, as we observe that our deterministic metapopulation-network modelling approach allows to evaluate the potential role of long range dispersal birds dispersal in the spatial spread of WNV in Germany. \par
To characterise the difference between the predictive ability of our  deterministic metapopulation and stochastic network model, we adhere to calculate the structural similarity (SSIM) index \cite{Massaro} for each case in each epidemiological seasons, with the suitable choices for $\phi$ and compare their distribution in Figure: \ref{Fig:SSIM}.
For more information please see the Supplementary Information for a detailed description.
 \begin{figure}[h!]
    \centering
     \includegraphics[width=0.5\textwidth]{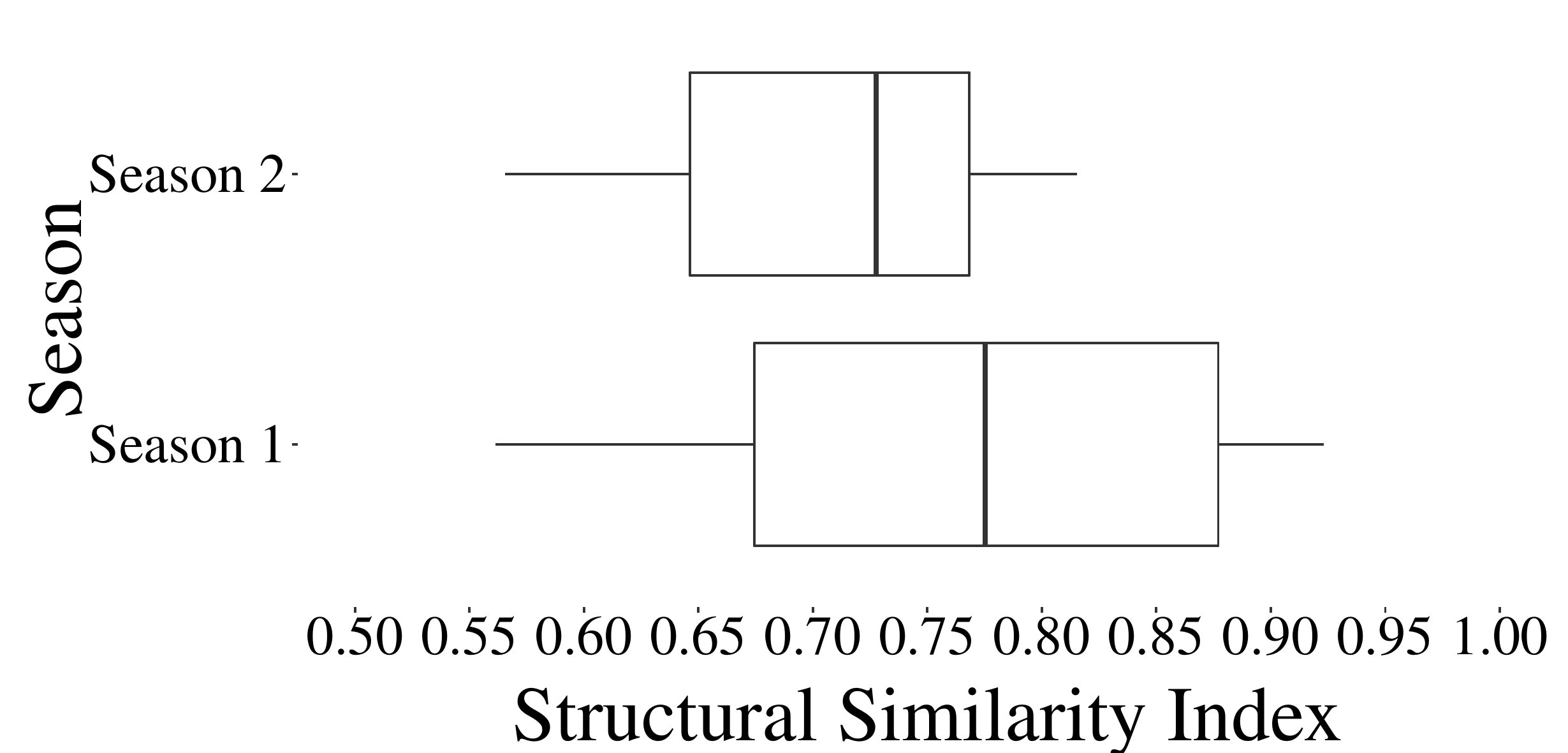}
    \caption{Spatial analysis of model outcome and the reported cases of WNV in Germany during the $2018-2019$ outbreaks. 
We compute SSIM score for each epidemiological season during the WNV outbreaks in Germany after $30$ simulations runs.
The distribution of SSIM scores or index are depicted as boxplots for each epidemiological season.
These boxplots can help us to understand the strength of our model while accounting the minimum, $1^\mathrm{st}$ quartile, median and $3^\mathrm{rd}$ quartile and the maximum of computed SSIM values.
It is noticeable  the range of SSIM scores in the first season is comparatively smaller to that off in the next season.}
    \label{Fig:SSIM}
\end{figure}

We can notice that in the initial phase of WNV spread in Germany, the model performs well while approximating the observed spatial dissemination.
In the following season though the performance of the model is slightly worse compared to the results from the previous season in the backdrop of reported cases.
Visual representations of the spatial $\mathcal {R}_{0}$ vs cumulative number of infected birds per Gemeinden are included in the Supplementary Information.

\section{Discussion \& Conclusion}

To comprehend the potential eco-epidemiological parameters and the disease mechanism yield to the preservation, appearance and ultimately lead to the potential spatial spread of WNV are essential to implement control strategies for the containment.
After the first case of WNV detected in Germany in 2018 \cite{ZIEGLER201939} and the favourable weather condition accentuated its possible spread across Germany especially not only confined to its first detected place in the  Eastern and Southeastern Germany. 
Following this sporadic cases of WNV, in the next season extraordinary high temperatures has allowed to decrease the extrinsic incubation period (EIP) and consequently the cases of WNV increased multifold \cite{Ute} across Germany. \par
In this work, we endeavour a deterministic metapopulation-network model of WNV transmission in Germany.
We put an effort to analyse the transmission of WNV locally and employ the contact networks to investigate the WNV transmission to the disease-free zones.
The contact networks incorporate different distinct types of vector and host movements: local bird movements in their habitat patch accordingly, long-distance movements of birds as well as vector movements.  
In our simulations we have opted for to include and to consider the different movement networks of the vector and the long range dispersal birds to capture the transmission patterns of WNV in Germany.
To procure the spatial prediction of WNV circulation across the whole Germany, we compare the spatial dissemination of WNV after including and afterwards ruling out the potential role of the long range dispersal bird's role in the spread of pathogens.
It is interesting to note that in the first season of WNV spread in Germany, the spatial transmission of WNV is well caught in a qualitative manner although in the next season similar feat is not being achieved after the inclusion of long range dispersal birds.
One possible reason could be attributed to the fact that the reported cases of WNV rely solely on the reporting and there might be reporting bias which might have hindered into the quantitative similarities of the reported cases and the simulated cases of WNV in Germany.
Another possible explanation could be the role long range dispersal birds what is benighted in the spread of WNV.
One thing should be noted that the uncertainty associated with the  number of overwintering infected mosquitoes can potentially trigger the outbreak.
From our simulations, it is clear that the role of long range dispersal birds to introduce the WNV in the different parts of Germany is recognisable and afterwards, the WNV circulation amongst the mosquito and the local birds help to localise the dissemination.
The spatial projections of the model outcomes can explain this.\par

Several previous models \cite{doi:10.1098/rspb.2017.1807, Tatem6242, 10.1371/journal.pcbi.1007184,Zhang2018BirdsMI, Cosner, LAPERRIERE201634}  include network models to analyse the role of birds and vectors in spatial disease transmission. 
In contrast to \cite{doi:10.1098/rspb.2017.1807, Tatem6242, 10.1371/journal.pcbi.1007184, Zhang2018BirdsMI, Cosner}, we included both, vector and host networks. 
In contrast to these studies, our results indicates that the role of vectors in the spatial spread of WNV cannot be neglected, even if information on vector movement is scarce and uncertainty is high. 
When reducing the vector population by $\approx 57 \%$, WNV dissemination can be controlled in our two patch model with the similar magnitude of the parameters.
Therefore, hypothetically one of the measure to consider to lower the WNV spread, is to control the vector population around the infected places.
\par
Our simulations have shown the potential role of the vector escalating and pursuing the vigorous course of WNV alive in the local scale and this might have an influence on the WNV transmission cycle.
Preceding \cite{MAIDANA2009403, Lewis, 1531-3492_2016_8_2423} spatial models try to investigate the dissemination of WNV utilising travelling wave solution with constant velocity but from the work of ~\cite{ZIEGLER201939, Ute}, it is evident that the transmission of WNV has resulted the cases of WNV across Germany from the North to the South~\cite{ZIEGLER201939, Ute}. 
Similar findings have been mentioned in \cite{10.1371/journal.pcbi.1006875, 1531-3492_2016_8_2423} of the features of the transmission of WNV in the USA.
According to such models, the reach of WNV dissemination should have been lesser compared to what we have noted in these years. 
The potential reason could be attributed to the fact that such models can not incorporate the spontaneous long distance movements of birds what are equally responsible to introduce the potential pathogens of WNV into the newer areas.
We surmise that long range dispersal birds movement network governed by the power-law model can potentially explain such phenomenon as we show in our current endeavour.
To our knowledge, the potential role of long range dispersal birds introducing WNV to the completely susceptible local bird population with the coupling of population dynamics with the host-vector mobility in Germany as we represent  in our theoretical modelling endeavour is not explored in other modelling studies.
\par
Considerable number of investigations on the spatial outbreaks of WNV have been performed.
The epidemic threshold condition for our model are in accordance with the with findings of \cite{Arin1}, although in our work we have taken account of the vector movement.
Our findings about the effectualness of the host-vector dispersal is similar to that in \cite{LAPERRIERE201634}.
The results from our mathematical model can theoretically mirror the fact that long range dispersal birds act as a discerning factor to introduce and transport WNV into new zones in Germany as mentioned in~\cite{ZIEGLER201939, Ute}.
\par
We opine that the power-law algorithm: \ref{Alg:MigBirdNet}, we use to generate the long range dispersal bird pathways in Germany sometimes have overvalued the potential cases of WNV as the algorithm constructs long-distance  fly pathways as depicted in Figure:\ref{Fig:Spatial} and the similar issues are mentioned in \cite{10.1371/journal.pcbi.1006875}.
Few of the employment of more realistic data driven dispersal functions as developed in \cite{Blasi, doi:10.1098/rsos.180438} can improve the current endeavour in a multifold while generating the migratory and the long range dispersal networks.
Knowledge about the factors like the density and the population distributions of the local bird populations, their habitat types, flying pathways will bring finer resolutions to the simulations results and the possible further incorporations of the different environmental parameters will yield better knowledge of the spread of WNV in Germany. 
\par
Reported cases of WNV amongst the human and the horses are outcome of a complex  patterns of biological, physical  and mechanical processes, which includes the virus transmission and population dynamics of  vector and amplifying hosts such as birds (including the local and the long range dispersal), effects of weather and the movements of the vector-hosts etc. 
This model allows us to explore the possible scenarios under which we are able to understand the possible spread of WNV in Germany and estimate the spatial spread.
This should potentially be helpful to identify the possible zones for the horse vaccination. 
Despite the limitations of our current effort on modelling the spatial spread of WNV quantitatively, qualitatively we are able to produce the path of infection. 
The projected WNV cases are similar to the observed reported cases except for some zones in the southern and the eastern zones.
This modelling investigation which can aid in the policy makers to induct the necessary assistance to the diverging needs across Germany.
Our modelling endeavours can be useful to simulate the project the potential cases of WNV in future and the output from our model can be useful to examine the strategies require to mitigate the cases of WNV in Germany.
Systematic applications of the approach proposed here will enable the future modelling endeavours to be more robust and with its necessary insights.

\bibliography{mybibfile}

\begin{thebibliography}{10}
\expandafter\ifx\csname url\endcsname\relax
  \def\url#1{\texttt{#1}}\fi
\expandafter\ifx\csname urlprefix\endcsname\relax\def\urlprefix{URL }\fi
\expandafter\ifx\csname href\endcsname\relax
  \def\href#1#2{#2} \def\path#1{#1}\fi

\bibitem{ZIEGLER201939}
U.~Ziegler, R.~L{\"u}hken, M.~Keller, D.~Cadar, E.~van~der Grinten, F.~Michel,
  K.~Albrecht, M.~Eiden, M.~Rinder, L.~Lachmann, D.~H{\"o}per,
  A.~Vina-Rodriguez, W.~Gaede, A.~Pohl, J.~Schmidt-Chanasit, M.~H. Groschup,
  West nile virus epizootic in germany, 2018, Antiviral Research 162 (2019) 39
  -- 43.

\bibitem{Chancey.2015}
C.~Chancey, A.~Grinev, E.~Volkova, M.~Rios, The global ecology and epidemiology
  of west nile virus, Biomed Res Int 2015 (2015) 376230.

\bibitem{doi:10.1111/1365-2435.12645}
K.~L. VanderWaal, V.~O. Ezenwa, Heterogeneity in pathogen transmission:
  mechanisms and methodology, Functional Ecology 30~(10) (2016) 1606--1622.

\bibitem{White7374}
L.~A. White, J.~D. Forester, M.~E. Craft, Disease outbreak thresholds emerge
  from interactions between movement behavior, landscape structure, and
  epidemiology, Proceedings of the National Academy of Sciences 115~(28) (2018)
  7374--7379.

\bibitem{10.1093/icb/icw015}
T.~Boulinier, S.~Kada, A.~Ponchon, M.~Dupraz, M.~Dietrich, A.~Gamble,
  V.~Bourret, O.~Duriez, R.~Bazire, J.~Tornos, T.~Tveraa, T.~Chambert,
  R.~Garnier, K.~D. McCoy, {Migration, Prospecting, Dispersal? What Host
  Movement Matters for Infectious Agent Circulation?}, Integrative and
  Comparative Biology 56~(2) (2016) 330--342.

\bibitem{10.1371/journal.pbio.2003489}
L.~L.~M. Shapiro, S.~A. Whitehead, M.~B. Thomas, Quantifying the effects of
  temperature on mosquito and parasite traits that determine the transmission
  potential of human malaria, PLOS Biology 15~(10) (2017) 1--21.

\bibitem{Kilpatrick2007}
A.~M. Kilpatrick, S.~L. LaDeau, P.~P. Marra, {Ecology of West Nile Virus
  Transmission and its Impact on Birds in the Western Hemisphere}, The Auk
  124~(4) (2007) 1121 -- 1136.

\bibitem{Rappole}
J.~H. Rappole, S.~R. Derrickson, Z.~Hub{\'a}lek, Migratory birds and spread of
  west nile virus in the western hemisphere, Emerging infectious diseases 6~(4)
  (2000) 319--328.

\bibitem{Reed}
K.~D. Reed, J.~K. Meece, J.~S. Henkel, S.~K. Shukla, Birds, migration and
  emerging zoonoses: west nile virus, lyme disease, influenza a and
  enteropathogens, Clinical medicine \& research 1~(1) (2003) 5--12.

\bibitem{newton2010migration}
I.~Newton, The migration ecology of birds, Elsevier, 2010.

\bibitem{https://doi.org/10.1111/ele.13335}
E.~A. Mordecai, J.~M. Caldwell, M.~K. Grossman, C.~A. Lippi, L.~R. Johnson,
  M.~Neira, J.~R. Rohr, S.~J. Ryan, V.~Savage, M.~S. Shocket, R.~Sippy, A.~M.
  Stewart~Ibarra, M.~B. Thomas, O.~Villena, Thermal biology of mosquito-borne
  disease, Ecology Letters 22~(10) (2019) 1690--1708.

\bibitem{10.1371/journal.pbio.3000938}
J.~R. Rohr, J.~M. Cohen, Understanding how temperature shifts could impact
  infectious disease, PLOS Biology 18~(11) (2020) 1--13.

\bibitem{rudolf2017west}
I.~Rudolf, L.~Bet{\'a}{\v{s}}ov{\'a}, H.~Bla{\v{z}}ejov{\'a},
  K.~Vencl{\'\i}kov{\'a}, P.~Strakov{\'a}, O.~{\v{S}}ebesta, J.~Mendel,
  T.~Bakonyi, F.~Schaffner, N.~Nowotny, et~al., West nile virus in
  overwintering mosquitoes, central europe, Parasites \& vectors 10~(1) (2017)
  1--4.

\bibitem{v12010123}
B.~Zana, K.~Erd{\'e}lyi, A.~Nagy, E.~Mezei, O.~Nagy, M.~Tak{\'a}cs, T.~Bakonyi,
  P.~Forg{\'a}ch, O.~Korbacska-Kutasi, O.~Feh{\'e}r, P.~Malik, K.~Ursu,
  P.~Kert{\'e}sz, A.~Kepner, M.~Martina, T.~S{\"u}li, Z.~Lanszki, G.~E.
  T{\'o}th, A.~Kuczmog, B.~Somogyi, F.~Jakab, G.~Kemenesi, Multi-approach
  investigation regarding the west nile virus situation in hungary, 2018,
  Viruses 12~(1) (2020).

\bibitem{v11070639}
V.~{\v C}abanov{\'a}, S.~{\v S}ikutov{\'a}, P.~Strakov{\'a}, O.~{\v S}ebesta,
  B.~Vichov{\'a}, D.~Zubr{\'\i}kov{\'a}, M.~Miterp{\'a}kov{\'a}, J.~Mendel,
  Z.~Hurn{\'\i}kov{\'a}, Z.~Hub{\'a}lek, I.~Rudolf, Co-circulation of west nile
  and usutu flaviviruses in mosquitoes in slovakia, 2018, Viruses 11~(7)
  (2019).

\bibitem{https://doi.org/10.1111/tbed.13452}
P.~de~Heus, J.~Kolodziejek, J.~V. Camp, K.~Dimmel, Z.~Bag{\'o}, Z.~Hub{\'a}lek,
  R.~van~den Hoven, J.-M.~V. Cavalleri, N.~Nowotny, Emergence of west nile
  virus lineage 2 in europe: Characteristics of the first seven cases of west
  nile neuroinvasive disease in horses in austria, Transboundary and Emerging
  Diseases 67~(3) (2020) 1189--1197.

\bibitem{doi:10.1098/rspb.2017.1807}
D.~R. Daversa, A.~Fenton, A.~I. Dell, T.~W.~J. Garner, A.~Manica, Infections on
  the move: how transient phases of host movement influence disease spread,
  Proceedings of the Royal Society B: Biological Sciences 284~(1869) (2017)
  20171807.

\bibitem{Tatem6242}
A.~J. Tatem, S.~I. Hay, D.~J. Rogers, Global traffic and disease vector
  dispersal, Proceedings of the National Academy of Sciences 103~(16) (2006)
  6242--6247.

\bibitem{Sumner}
T.~Sumner, R.~J. Orton, D.~M. Green, R.~R. Kao, S.~Gubbins, Quantifying the
  roles of host movement and vector dispersal in the transmission of
  vector-borne diseases of livestock, PLoS computational biology 13~(4) (2017)
  e1005470--e1005470.

\bibitem{Ute}
U.~Ziegler, P.~D. Santos, M.~H. Groschup, C.~Hattendorf, M.~Eiden,
  D.~H{\"o}per, P.~Eisermann, M.~Keller, F.~Michel, R.~Klopfleisch,
  K.~M{\"u}ller, D.~Werner, H.~Kampen, M.~Beer, C.~Frank, R.~Lachmann, B.~A.
  Tews, C.~Wylezich, M.~Rinder, L.~Lachmann, T.~Gr{\"u}newald, C.~A. Szentiks,
  M.~Sieg, J.~Schmidt-Chanasit, D.~Cadar, R.~L{\"u}hken, West nile virus
  epidemic in germany triggered by epizootic emergence, 2019, Viruses 12~(4)
  (2020) 448.

\bibitem{MAIDANA2009403}
N.~A. Maidana, H.~M. Yang, Spatial spreading of west nile virus described by
  traveling waves, Journal of Theoretical Biology 258~(3) (2009) 403 -- 417.

\bibitem{Tar}
A.~K. Tarboush, Z.~Lin, M.~Zhang, Spreading and vanishing in a west nile virus
  model with expanding fronts, Science China Mathematics 60~(5) (2017)
  841--860.

\bibitem{Wang}
Z.~Wang, H.~Nie, Y.~Du, Spreading speed for a west nile virus model with free
  boundary, Journal of Mathematical Biology 79~(2) (2019) 433--466.

\bibitem{kenkre2005theoretical}
V.~Kenkre, R.~R. Parmenter, I.~Peixoto, L.~Sadasiv, A theoretical framework for
  the analysis of the west nile virus epidemic, Mathematical and computer
  modelling 42~(3-4) (2005) 313--324.

\bibitem{BICHARA2016128}
D.~Bichara, C.~Castillo-Chavez, Vector-borne diseases models with residence
  times -- a lagrangian perspective, Mathematical Biosciences 281 (2016) 128 --
  138.

\bibitem{COLIZZA2008450}
V.~Colizza, A.~Vespignani, Epidemic modeling in metapopulation systems with
  heterogeneous coupling pattern: Theory and simulations, Journal of
  Theoretical Biology 251~(3) (2008) 450 -- 467.

\bibitem{Arino06diseasespread}
J.~Arino, P.~V.~D. Driessche, Disease spread in metapopulations, Fields Inst.
  Commun (2006) 1--12.

\bibitem{doi:10.1142/9789814261265_0003}
J.~Arino, Diseases in Metapopulations, pp. 64--122.

\bibitem{Arin1}
J.~Arino, A.~Ducrot, P.~Zongo, A metapopulation model for malaria with
  transmission-blocking partial immunity in hosts, Journal of Mathematical
  Biology 64~(3) (2012) 423--448.

\bibitem{Arin2}
J.~Arino, Spatio-temporal spread of infectious pathogens of humans, Infectious
  Disease Modelling 2~(2) (2017) 218--228.

\bibitem{refId0}
{Durand, Benoit}, {Balan\c{c}a, Gilles}, {Baldet, Thierry}, {Chevalier,
  V\'eronique}, A metapopulation model to simulate west nile virus circulation
  in western africa, southern europe and the mediterranean basin, Vet. Res.
  41~(3) (2010) 32.

\bibitem{doi:10.1137/18M1236162}
F.-B. Wang, R.~Wu, X.-Q. Zhao, A west nile virus transmission model with
  periodic incubation periods, SIAM Journal on Applied Dynamical Systems 18~(3)
  (2019) 1498--1535.

\bibitem{Gumel}
C.~Bowman, A.~B. Gumel, P.~van~den Driessche, J.~Wu, H.~Zhu, A mathematical
  model for assessing control strategies against west nile virus, Bulletin of
  Mathematical Biology 67~(5) (2005) 1107--1133.

\bibitem{BHOWMICK2020110117}
S.~Bhowmick, J.~Gethmann, F.~J. Conraths, I.~M. Sokolov, H.~H. Lentz, Locally
  temperature - driven mathematical model of west nile virus spread in germany,
  Journal of Theoretical Biology 488 (2020) 110117.

\bibitem{10.1371/journal.pcbi.1006875}
S.~A. Moon, L.~W. Cohnstaedt, D.~S. McVey, C.~M. Scoglio, A spatio-temporal
  individual-based network framework for west nile virus in the usa: Spreading
  pattern of west nile virus, PLOS Computational Biology 15~(3) (2019) 1--24.

\bibitem{refId1}
{Durand, Benoit}, {Balan\c{c}a, Gilles}, {Baldet, Thierry}, {Chevalier,
  V\'eronique}, A metapopulation model to simulate west nile virus circulation
  in western africa, southern europe and the mediterranean basin, Vet. Res.
  41~(3) (2010) 32.

\bibitem{LAPERRIERE201634}
V.~Laperri{\`e}re, K.~Brugger, F.~Rubel, Cross-scale modeling of a vector-borne
  disease, from the individual to the metapopulation: The seasonal dynamics of
  sylvatic plague in kazakhstan, Ecological Modelling 342 (2016) 34 -- 48.

\bibitem{doi:10.1086/521911}
D.~J. Gubler, The continuing spread of west nile virus in the western
  hemisphere, Clinical Infectious Diseases 45~(8) (2007) 1039--1046.

\bibitem{Styer2007}
L.~M. Styer, K.~A. Kent, R.~G. Albright, C.~J. Bennett, L.~D. Kramer, K.~A.
  Bernard, Mosquitoes inoculate high doses of west nile virus as they probe and
  feed on live hosts, PLoS Pathog 3~(9) (2007) 1262--1270.

\bibitem{Vogels2017}
C.~B. Vogels, G.~P. G{\"o}ertz, G.~P. Pijlman, C.~J. Koenraadt, Vector
  competence of european mosquitoes for west nile virus, Emerging Microbes
  \&Amp; Infections 6 (2017) e96 EP --.

\bibitem{10.1371/journal.pntd.0002768}
G.~L. Hamer, T.~K. Anderson, D.~J. Donovan, J.~D. Brawn, B.~L. Krebs, A.~M.
  Gardner, M.~O. Ruiz, W.~M. Brown, U.~D. Kitron, C.~M. Newman, T.~L. Goldberg,
  E.~D. Walker, Dispersal of adult culex mosquitoes in an urban west nile virus
  hotspot: A mark-capture study incorporating stable isotope enrichment of
  natural larval habitats, PLOS Neglected Tropical Diseases 8~(3) (2014) 1--7.

\bibitem{doi:10.1093/jmedent/31.3.508}
E.~A. Myhre, R.~D. Akre, What bit me? identifying hawaii's stinging and biting
  insects and their kin, Journal of Medical Entomology 31~(3) (1994) 508.

\bibitem{VERDONSCHOT201469}
P.~F. Verdonschot, A.~A. Besse-Lototskaya, Flight distance of mosquitoes
  (culicidae): A metadata analysis to support the management of barrier zones
  around rewetted and newly constructed wetlands, Limnologica 45 (2014) 69 --
  79.

\bibitem{Cluexvol}
E.~Vinogradova, Culex Pipiens Pipiens Mosquitoes: Taxonomy, Distribution,
  Ecology, Physiology, Genetics, Applied Importance and Control, no. ISBN
  978-9546421036 in Pensoft Series Parasitologica, no. 2., Sofia : Pensoft,
  2000, 2000.

\bibitem{doi:10.1002/eap.1612}
Y.~Alcalay, I.~Tsurim, O.~Ovadia, Modelling the effects of spatial
  heterogeneity and temporal variation in extinction probability on mosquito
  populations, Ecological Applications 27~(8) (2017) 2342--2358.

\bibitem{MOULAY2013129}
D.~Moulay, Y.~Pign{\'e}, A metapopulation model for chikungunya including
  populations mobility on a large-scale network, Journal of Theoretical Biology
  318 (2013) 129 -- 139.

\bibitem{campbell2002west}
G.~L. Campbell, A.~A. Marfin, R.~S. Lanciotti, D.~J. Gubler, West nile virus,
  The Lancet infectious diseases 2~(9) (2002) 519--529.

\bibitem{liu2006modeling}
R.~Liu, J.~Shuai, J.~Wu, H.~Zhu, Modeling spatial spread of west nile virus and
  impact of directional dispersal of birds, Mathematical Biosciences \&
  Engineering 3~(1) (2006) 145.

\bibitem{hartemink2007importance}
N.~Hartemink, S.~Davis, P.~Reiter, Z.~Hub{\'a}lek, J.~Heesterbeek, Importance
  of bird-to-bird transmission for the establishment of west nile virus,
  Vector-Borne and Zoonotic Diseases 7~(4) (2007) 575--584.

\bibitem{8937990}
M.~Beal, (pat)terns in space and time: Movement, activity, and habitat
  preference in breeding caspian terns (hydroprogne caspia) (2018).

\bibitem{Butler2018}
S.~R. Butler, J.~J. Templeton, E.~Fern{\'a}ndez-Juricic, How do birds look at
  their world? a novel avian visual fixation strategy, Behavioral Ecology and
  Sociobiology 72~(3) (2018) 38.

\bibitem{ref2018-32206-003}
S.~E. Cohen, P.~M. Todd, Relationship foraging: Does time spent searching
  predict relationship length?, Evolutionary Behavioral Sciences 12~(3) (2018)
  139--151.

\bibitem{https://doi.org/10.1111/1365-2745.12685}
C.~Garc{\'\i}a, L.~Borda-de {\'A}gua, Extended dispersal kernels in a changing
  world: insights from statistics of extremes, Journal of Ecology 105~(1)
  (2017) 63--74.

\bibitem{10.1371/journal.pone.0156688}
N.~S. Da~Silveira, B.~B.~S. Niebuhr, R.~d.~L. Muylaert, M.~C. Ribeiro, M.~A.
  Pizo, Effects of land cover on the movement of frugivorous birds in a
  heterogeneous landscape, PLOS ONE 11~(6) (2016) 1--19.

\bibitem{trove.nla.gov.au/work/217129952}
I.~Donoso, D.~Garcia, J.~Rodriguez-Perez, D.~Martinez, Incorporating seed fate
  into plant-frugivore networks increases interaction diversity across plant
  regeneration stages.(report), Oikos 125~(12) (2016-12-01) 1762(10).

\bibitem{doi:10.1111/j.1365-2745.2008.01401.x}
D.~J. Levey, J.~J. Tewksbury, B.~M. Bolker, Modelling long-distance seed
  dispersal in heterogeneous landscapes, Journal of Ecology 96~(4) (2008)
  599--608.

\bibitem{Banos-Villalba2017}
A.~B{\'a}{\~n}os-Villalba, G.~Blanco, J.~A. D{\'a}z-Luque, F.~V. D{\'e}nes,
  F.~Hiraldo, J.~L. Tella, Seed dispersal by macaws shapes the landscape of an
  amazonian ecosystem, Scientific Reports 7~(1) (2017) 7373.

\bibitem{doi:10.1111/j.1365-2745.2008.01379.x}
T.~A. Carlo, J.~M. Morales, Inequalities in fruit-removal and seed dispersal:
  consequences of bird behaviour, neighbourhood density and landscape
  aggregation, Journal of Ecology 96~(4) (2008) 609--618.

\bibitem{NATHAN2008638}
R.~Nathan, F.~M. Schurr, O.~Spiegel, O.~Steinitz, A.~Trakhtenbrot, A.~Tsoar,
  Mechanisms of long-distance seed dispersal, Trends in Ecology \& Evolution
  23~(11) (2008) 638 -- 647.

\bibitem{doi:10.1111/j.1469-8137.2011.04051.x}
M.~Nogales, R.~Heleno, A.~Traveset, P.~Vargas, Evidence for overlooked
  mechanisms of long-distance seed dispersal to and between oceanic islands,
  New Phytologist 194~(2) (2012) 313--317.

\bibitem{doi:10.1111/j.0906-7590.2006.04677.x}
M.~Calvi{\~n}o-Cancela, R.~R.~Dunn, E.~J.~B. Van~Etten, B.~B.~Lamont, Emus as
  non-standard seed dispersers and their potential for long-distance dispersal,
  Ecography 29~(4) (2006) 632--640.

\bibitem{doi:10.1890/15-0734.1}
J.~D. Herrmann, T.~A. Carlo, L.~A. Brudvig, E.~I. Damschen, N.~M. Haddad, D.~J.
  Levey, J.~L. Orrock, J.~J. Tewksbury, Connectivity from a different
  perspective: comparing seed dispersal kernels in connected vs. unfragmented
  landscapes, Ecology 97~(5) (2016) 1274--1282.

\bibitem{10.1371/journal.pone.0193660}
{\'O}.~M. ~, J.~C. Bicca-Marques, C.~A. Chapman, Quantity and quality of seed
  dispersal by a large arboreal frugivore in small and large atlantic forest
  fragments, PLOS ONE 13~(3) (2018) 1--16.

\bibitem{SeedDispersalMapper}
B.~B.~S. Niebuhr, F.~S. Pinto, J.~W. Ribeiro, K.~M. Costa, R.~F.~B. Silva,
  M.~C. Ribeiro, Predicting natural regeneration through landscape structure,
  movement of frugivore fauna, and seed dispersal.

\bibitem{10.1093/aobpla/plz042}
H.~S. Rogers, N.~G. Beckman, F.~Hartig, J.~S. Johnson, G.~Pufal, K.~Shea,
  D.~Zurell, J.~M. Bullock, R.~S. Cantrell, B.~Loiselle, L.~Pejchar, O.~H.
  Razafindratsima, M.~E. Sandor, E.~W. Schupp, W.~C. Strickland, J.~Zambrano,
  {The total dispersal kernel: a review and future directions}, AoB PLANTS
  11~(5) (09 2019).

\bibitem{https://doi.org/10.1111/j.1365-2745.2008.01401.x}
D.~J. Levey, J.~J. Tewksbury, B.~M. Bolker, Modelling long-distance seed
  dispersal in heterogeneous landscapes, Journal of Ecology 96~(4) (2008)
  599--608.

\bibitem{doi:10.1080/00107510500052444}
M.~Newman, Power laws, pareto distributions and zipf's law, Contemporary
  Physics 46~(5) (2005) 323--351.

\bibitem{meyer2014}
S.~Meyer, L.~Held, Power-law models for infectious disease spread, Ann. Appl.
  Stat. 8~(3) (2014) 1612--1639.

\bibitem{Abraham}
A.~Berman, R.~J. Plemmons, Chapter 2 - nonnegative matrices, in: A.~Berman,
  R.~J. Plemmons (Eds.), Nonnegative Matrices in the Mathematical Sciences,
  Academic Press, 1979, pp. 26 -- 62.

\bibitem{VANDENDRIESSCHE200229}
P.~van~den Driessche, J.~Watmough, Reproduction numbers and sub-threshold
  endemic equilibria for compartmental models of disease transmission,
  Mathematical Biosciences 180~(1) (2002) 29 -- 48.

\bibitem{Ciota}
A.~T. Ciota, L.~D. Kramer, Vector-virus interactions and transmission dynamics
  of west nile virus, Viruses 5~(12) (2013) 3021--3047.

\bibitem{RUAN201765}
S.~Ruan, Modeling the transmission dynamics and control of rabies in china,
  Mathematical Biosciences 286 (2017) 65 -- 93.

\bibitem{dCode}
dcode, \url{https://www.dcode.fr/en}, accessed: 2019-09-30.

\bibitem{Massaro}
E.~Massaro, D.~Kondor, C.~Ratti, Assessing the interplay between human mobility
  and mosquito borne diseases in urban environments, Scientific Reports 9~(1)
  (2019) 16911.

\bibitem{Lewis2019}
M.~A. Lewis, Z.~Shuai, P.~van~den Driessche, A general theory for target
  reproduction numbers with applications to ecology and epidemiology, Journal
  of Mathematical Biology 78~(7) (2019) 2317--2339.

\bibitem{TierSeuchenInformationsSystem}
Tierseucheninformationssystem,
  \url{https://www.fli.de/de/presse/pressemitteilungen/presse-einzelansicht/fli-stellt-erstmals-west-nil-virus-infektion-bei-einem-vogel-in-deutschland-fest/},
  accessed: 2020-09-30.

\bibitem{10.1371/journal.pcbi.1007184}
L.~W. Pomeroy, H.~Kim, N.~Xiao, M.~Moritz, R.~Garabed, Network analyses to
  quantify effects of host movement in multilevel disease transmission models
  using foot and mouth disease in cameroon as a case study, PLOS Computational
  Biology 15~(8) (2019) 1--17.

\bibitem{Zhang2018BirdsMI}
J.~Zhang, Z.~Jin, H.~Zhu, Birds movement impact on the transmission ofwest nile
  virus between patches, Journal of Applied Analysis and Computation 8 (2018)
  443--456.

\bibitem{Cosner}
J.~Zhang, C.~Cosner, H.~Zhu, Two-patch model for the spread of west nile virus,
  Bulletin of Mathematical Biology 80~(4) (2018) 840--863.

\bibitem{Lewis}
M.~Lewis, J.~Renc{\l}awowicz, P.~v. den Driessche, Traveling waves and spread
  rates for a west nile virus model, Bulletin of Mathematical Biology 68~(1)
  (2006) 3--23.

\bibitem{1531-3492_2016_8_2423}
J.~Chen, J.~Huang, J.~C. Beier, R.~S. Cantrell, C.~Cosner, D.~O. Fuller,
  G.~Zhang, S.~Ruan, Modeling and control of local outbreaks of west nile virus
  in the united states (2016).

\bibitem{Blasi}
A.~K{\"o}lzsch, B.~Blasius, Theoretical approaches to bird migration, The
  European Physical Journal Special Topics 157~(1) (2008) 191--208.

\bibitem{doi:10.1098/rsos.180438}
A.~K{\"o}lzsch, E.~Kleyheeg, H.~Kruckenberg, M.~Kaatz, B.~Blasius, A periodic
  markov model to formalize animal migration on a network, Royal Society Open
  Science 5~(6) (2018) 180438.

\end{thebibliography}

\end{document}